\documentclass[preprint2]{aastex}

\newcommand{\rrl}{{RR~Lyrae}}

\usepackage{longtable}
\slugcomment{Not to appear in Nonlearned J., 45.}
\shorttitle{Variable Stars in NGC~2419}
\shortauthors{Di Criscienzo et al.}
\begin{document}
\title{A New Census of the variable star population in the Globular Cluster \\ NGC~2419\thanks{Based on data collected at the 3.5 m Telescopio Nazionale
Galileo, operated by INAF, and the SUBARU and Hubble Space Telescope archives}}

\author{M. Di Criscienzo\altaffilmark{1}, C. Greco\altaffilmark{2,3}, V. Ripepi\altaffilmark{4}, G. Clementini\altaffilmark{3}, 
M. Dall' Ora\altaffilmark{4}, M. Marconi\altaffilmark{4}, I. Musella\altaffilmark{4}, L. Federici\altaffilmark{3}, L. Di Fabrizio\altaffilmark{5}}
\affil{$^{1}$INAF- Osservatorio Astronomico di Roma, via Frascati, 33, Monte Porzio Catone, I-00040, Rome, Italy}
%\email{dicrisci@na.astro.it}
%\author{{\it et al}.}
%\author{C. Greco}
\affil{$^{2}$Observatoire de Geneve, 51, ch. Des Maillettes, CH-1290 Sauverny, Switzerland}%; claudia.greco@unige.ch}
\affil{$^{3}$INAF- Osservatorio Astronomico di Bologna, via Ranzani 1, I-40127, Bologna Italy}
%\author{V. Ripepi}
\affil{$^{4}$INAF- Osservatorio Astronomico di Capodimonte, via Moiarello 16, I-80131 Napoli, Italy}
\affil{$^{5}$INAF-Centro Galileo Galilei \& Telescopio Nazionale Galileo, PO Box 565, 38700 S. Cruz de La Palma, Spain}

\begin{abstract}
We present $B, V$ and $I$ CCD light curves for 101 variable stars belonging to the globular cluster NGC~2419, 60 of which are new discoveries,
based on datasets obtained at the TNG, SUBARU and HST telescopes. The sample includes 75  RR Lyrae stars (of which 38 RRab, 36 RRc and one RRd), 
one  Population II Cepheid, 12 SX Phoenicis variables, 2 $\delta$Scuti stars, 3 binary systems, 5 long-period variables, and 3 variables of 
uncertain classification. 
The pulsation properties of the RR Lyrae variables  are close to those of
Oosterhoff type II clusters, consistent with the low metal abundance and the cluster
horizontal branch morphology, disfavoring (but not totally ruling out) an
extragalactic hypotesis for the origin of NGC~2419. The observed properties of RR Lyrae and SX Phoenicis stars are used to estimate the cluster reddening and distance,
 using a number of different methods. Our final value is  $\mu_0$ (NGC~2419) = 19.71 $\pm$ 0.08 mag (D= 87.5 $\pm$ 3.3 kpc), with $E(B-V)$= 0.08 $\pm$ 0.01 mag, 
[Fe/H]=$-$2.1 dex in the Zinn \& West metallicity scale, and a value of M$_V$ that sets $\mu_0$ (LMC) = 18.52 mag. This value is in good agreement with most recent literature estimates 
of the distance to NGC~2419. %The origin of this cluster is still unknown.

\end{abstract}


\section{Introduction}

Galactic Globular Clusters (GGCs) offer a fundamental tool for both
tracing the properties of our Galaxy in the present epoch, and for
piecing together its formation history.  They are old, massive, spread
all over the Galaxy and generally bright, hence can be observed up to
large Galactocentric distances ($ \simeq 100$ kpc) and through regions
of high extinction.  GGCs have long been used as tools to understand
how the Galactic halo formed. Two alternative scenarios have been
suggested in the past: according to Eggen, Sandage \& Lynden Bell (1962) the halo formed by
the rapid dissipative collapse of a single massive protogalactic
cloud, whereas \citet{sz78} suggested it was assembled over an
extended time via infall and capture of smaller fragments.  Currently,
the most favourite scenario implies that the main body of the Milky
Way (MW, $R_{\rm{gc}} < 10$ kpc) formed in a monolithic collapse,
while the outer parts ($R_{\rm{gc}} > 15$ kpc) were assembled by capture
of protogalactic fragments or at least experienced a number of merging
events (\citealt{vdbm04}). According to this scenario a fraction of
the ``young halo'' clusters is expected to have formed in external
dwarf galaxies, and to have later accreted when the parent galaxies
merged and were destroyed by the MW.  Many signatures of past and
present interactions have been found in the MW halo and a number of
GGCs may have an extragalactic origin. This is the case of
$\omega$ Cen and M54.  These clusters are both ``peculiar'' in some
respects.  $\omega$ Cen hosts multiple stellar populations (see
e.g. \citealt{bedin}; \citealt{rey}; \citealt{sollima}) and is
suggested to be the stripped core of a disrupted dwarf galaxy
(\citealt{villanova07}, and references therein). On the other hand,
M54 is thought to belong to the Sagittarius (Sgr) dwarf spheroidal
galaxy (dSph) that is currently merging with the MW (see
e.g. \citealt{layden}, \citealt{Monaco}, \citealt{be08}).
There is another GC in the Milky Way for which an extragalactic
origin was suggested: NGC2419, one of the most remote and luminous cluster in the MW.  
As $\omega$ Cen and M54, NGC2419 is significantly looser than expected
for {\bf its} brightness in the $M_V$ {\it vs} half-light radius ($R_h$)
plane (\citealt{mackey05}).\\  Apart from the similarity with $\omega$ Cen
and M54, other two pieces of evidence support the idea that
NGC~2419 might have an extragalactic origin and be the remnant of a
dwarf galaxy tidally disrupted by the MW (\citealt{vdbm04},
\citealt{mackey05}):
\begin{enumerate}
\item NGC~2419 has a large core radius, $r_c =0.32^{\prime}$
  \citep{be07}, corresponding to $r_c \sim 8$ pc at its distance \citep[$\sim
  90$ kpc,][]{ripepi07}, and the cluster half-light radius \citep[$r_h =0.96^{\prime}$,][]{be07}, corresponding to $r_h \sim $
  23 pc, is much larger than that of GCs of similar luminosity, and
  about as large as the largest nuclei of dwarf elliptical galaxies
  (\citealt{mackey05});
\item \citet{newberg}
find that NGC~2419 appears to lie within an overdensity of
A-type stars connected to
the 
tidal tails of the Sagittarius dSph, and suggest that the cluster might once
be associated to this galaxy.
\end{enumerate}
NGC~2419 has a central velocity dispersion
\citep[$\sigma_0$=4.14$\pm$0.48 km/s$^{-1}$,][]{baumgard} lower than
that of other dSphs for which this quantity was measured. According to 
this value in the $L_V$ {\it vs} $\sigma_0$ plane (\citealt{faber}), 
the cluster lies {\bf 3 $\sigma$} below the ``fundamental plane" relations for
Galactic GCs (see \citealt{degrjis}). NGC~2419 has other unusual properties 
as a globular cluster NGC~2419 has unusual properties also as a globular cluster.  

Its Galactocentric distance ($d = 90$ kpc) is typical of an outer halo
cluster, however the cluster is definitely metal poorer ([Fe/H] = $-$2.10
dex, \citealt{zw84}, \citealt{suntzeff}, [Fe/H] = $-$2.20 $\pm$0.009
dex, \citealt{carretta2009c}) than other outer halo clusters and
belongs to the most metal-poor group of MW GCs most of which (with the
only exception of AM-4) are located within $R_{\rm GC}$ $\simeq$ 20
Kpc. NGC~2419 is also one of the five most luminous GCs in the MW
(M$_{V}$= $-$9.58 mag, \citealt{harris96}, $-$9.4 mag,
\citealt{be07,baumgard})  breaking the rule that GCs in the outer 
halo (and those at  $R_{\rm{GC} } \ge 50$ in particular) are, on
average, fainter than GCs in the inner halo (\citealt{vdbm04}).\\
In conclusion, NGC~2419 appears to be peculiar both in terms of being
a GC (either Galactic or external) and as the possible remnant of a disrupted dwarf galaxy.\\

After the pionering photographic study of \citet{rh75} who published a
first color-magnitude-diagram (CMD) of the cluster deep to $V$ =22.2
mag, NGC~2419 has been the subject of many photometric studies both
from the ground and with the Hubble Space Telescope (HST).
CMDs for the cluster were published by \citet{christian}, %ground-based CCD photometry to $V \sim$ =24.0 mag;
\citet{harris97}, %Wide Field Planetary Camera 2 (WFPC2) on board of the HST photometry to $V \sim$ =27.8.0 mag;
\citet{stetson98,stetson05,saha}, 
%ground-based and WFPC2@HST photometry to $V \sim$ =23.5 mag;
\citet{sirianni}, 
% Advanced Camera for Surveys (ACS)
%on board of the HST photometry to $V \sim$ =26.3 mag;
and, more recently, by \citet{be07}, 
%, ACS photometry to $V \sim$ =24.5-25 mag;
\citet{ripepi07},  %, ground-based photometry to $V \sim$ =25.5 mag;
\citet{dalessandro}, and %, ACS photometry to $V \sim$ =26.2 mag;
\citet{sandquist}.%, ACS photometry to $V \sim$ =26.2 mag. 
 Due to the large distance, the ground-based CMDs of NGC~2419 do not
generally go fainter than the cluster main sequence turn-off (TO).
On the other hand, the HST-based CMDs, while deeper,
(\citealt{harris97}'s in particular, which reaches $V \sim$ 27.8 mag),
cover only small portions of the cluster and, also, generally lack
stars brighter than $V \sim$ 18-19 mag, since often they are
saturated in the HST images. \\
In \citet{ripepi07} we published a CMD of NGC~2419 which, for the
first time, is both deep and covers a very large field of view around
the cluster. 
Our CMD samples a total range of about 9 magnitudes, from the tip of the cluster red giant branch  
(RGB) around $V \sim$ 17 mag, down to $V \sim$ 25.7 mag, about  2.3
mag below the TO, and extends over an area encompassing more than 1
tidal radius ($r_t=8.74^{\prime}$, \citealt{trager}) in the
North-South direction and about 2 tidal radii in East-West from the 
cluster center (see Fig.\ref{campo}).  
\citet{ripepi07} CMD shows that the horizontal branch (HB) of NGC~2419 extends
down to an extremely long blue tail ending with the 
``blue hook'' for $V \ge 23.8$ mag \citep[see][for a detailed
discussion of the blue hook origin in NGC2419 and other GGCs]{brown10}. The cluster also
has a prominent Blue Straggler Star (BSS)
sequence populated by an extraordinary large number of BSSs \citep[see also][]{dalessandro}.\\
As anticipated in \citet{ripepi07},  
we have performed a new extensive study of the cluster variable stars and
detected 60 new variables, among which 39 are RR Lyrae stars that
double the cluster RR Lyrae population, and 12 are SX Phoenicis (SX Phe) 
variables located in the BSS region of the CMD. The only previous study of the 
NGC~2419 variables is by Pinto \& Rosino (1977, hereafter PR) and is based on photographic plates and
``by eye'' photometry. PR identified 41 variables mainly located in
the outer parts of the cluster, of which 32 were recognized as RR
Lyrae stars.  In this paper we present a full description of the
database we employed in \citet{ripepi07}, describe in detail the study
and properties of the variable stars and publish the catalogue of
light curves.  Observations and data reduction techniques used to
process the total time-series of NGC~2419 are described in Section 2.
In Section 3 we detail the methods used to detect and study the
variable stars, and present the complete catalog of light curves.  In
Section 4 we describe the analysis of the various types of variables
(RR Lyrae, SX Phe, long period and miscellaneous variables) and their
distribution on the cluster CMD.  In Section 5 we discuss the cluster
distance based on its RR Lyrae and SX Phe stars. In particular, we
compare the observed properties of the RR Lyrae stars with the
predictions of nonlinear pulsation models to constrain the cluster
distance modulus. Finally, in Section 6 we provide a summary of our
results.

\section{Observations and data reduction}
\begin{figure}[h]
\begin{center}
\includegraphics[width=8cm]{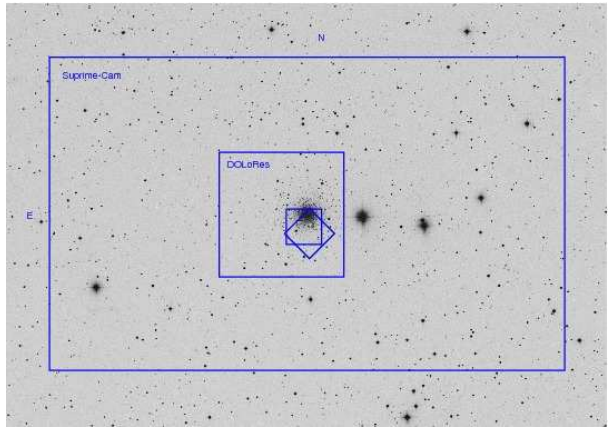}
\caption{Total field of view covered by our data for NGC~2419. Boxes mark the fields covered by each of the 3 different imagers we used in our study: 
Suprime-CAM at the SUBARU telescope (largest box), DOLORES at the TNG, and WFPC2 on board the HST (smallest boxes).}
\label{campo}
\end{center}
\end{figure}

Time-series $B, V$ photometric observations of NGC~2419
(RA=07h38m24.0s, decl=38$^o$454$'$00'', J2000) were obtained between
2003 September and 2004 February with DOLORES at the 3.5m TNG
telescope\footnote{http://www.tng.iac.es/instruments/lrs/}.
Observations were performed in service mode under the requirement of
seeing better than 1.5 arcsec.  The TNG data were complemented by
F555W and F814W Wide Field Planetary Camera 2 (WFPC2) on board of the
HST\footnote{http://www.stsci.edu/hst/wfpc2} archival observations of
the cluster spanning 6 years from 1994, May to 2000, March (Programs
ID\ 5481, 7628, 7630, 8095), and by $V, I$ photometry obtained with the
Suprime-Cam of the SUBARU 8.2 m
telescope\footnote{http://www.subarutelescope.org/Observing/Instruments/SCam/index.html} along four nights in 2002, December. Typical
exposure times of the HST observations range from 40s to 1400s in F555W, and from 40s to 1300s in F814W. The
SUBARU data consist instead of 30s and 180s exposures in  both the $I$ and
$V$ bands.  The total time-series dataset comprises 20 $B$, 212 $V$
and 65 $I$ images.  Logs of the observations and details of the
instrumental set-up at the various telescopes are provided in Table
\ref{tobs}.  The large field of view of the Suprime-Cam (34 $\times$
27 arcmin$^2$), covered by a mosaic of 10 CCDs, and dithering of the
telescope pointings resulted in the survey of a total area of 50
$\times$ 43 arcmin$^2$ centered on NGC~2419, which includes both the
TNG and HST fields.
Results presented in this paper cover a region extending $\pm
10.5^{\prime}$ in North-South and $\pm 18.3^{\prime}$ in East-West from the cluster
center. 
The area, which corresponds to five  partially overlapping CCDs of the 
SUBARU mosaic, is shown in Fig. \ref{campo}.\\
Images were pre-reduced following standard techniques (bias
subtraction and flat-field correction) with IRAF\footnote{IRAF is
  distributed by the National Optical Astronomical Observatories,
  which are operated by the Association of Universities for Research
  in Astronomy, Inc., under cooperative agreement with the National
  Science Foundation}.  We then performed PSF photometry of the
pre-processed images running the DAOPHOTII/ALLSTAR/ALLFRAME packages
\citep{stetson87,stetson92} on the TNG, HST and SUBARU datasets,
separately.  Typical internal errors of the $V$ band photometry per
single phase point at the level of the HB are in the range from 0.01
to 0.15 mag. We checked for consistency the photometry by randomly
extracting several pairs of images and computing the differences of
the magnitudes of the stars in common to each pair, which showed a
scatter consistent with the intrinsic photometric error, with no trend
with magnitude and position
of the stars along the frames.\\
The absolute photometric calibration was obtained by using local
standards in NGC~2419 from P.B.  Stetson's list\footnote{Available at
  http://cadcwww.dao.nrc.ca/standards/}.  A total of about 400
standard stars covering approximately the color intervals $-0.2 <
(B-V) < 1.8$ mag and $-0.2 < (V-I) < 2.0$ mag were used to derive a
set of linear calibration equations for $B$, $V$ and $I$,
respectively:  
\begin{center}
\begin{eqnarray*}
B&=&b_{TNG}+(0.017\pm0.003)\times(b_{TNG}-v_{SUB})+\\&&+(7.709\pm0.0003)\\
V&=&v_{SUB}+(6.123\pm 0.014) \\
I&=&i_{SUB}+(0.039\pm 0.002)\times(v_{SUB}-i_{SUB})+\\ &&+(5.811 \pm 0.002)
\end{eqnarray*}
\end{center}
\noindent
Zero point uncertainties
are of 0.022, 0.014 and 0.014 mag
in $B$, $V$ and $I$, respectively. \\ The $V$, $V-I$ CMD of NGC~2419 based on the SUBARU dataset was published  and discussed in \citet{ripepi07}.
%and will be further investigated in a forthcoming paper. 
Here we focus on the cluster variable stars.

\section[]{Identification of the variable stars}
\begin{figure*}
\begin{center}
\includegraphics[width=10cm]{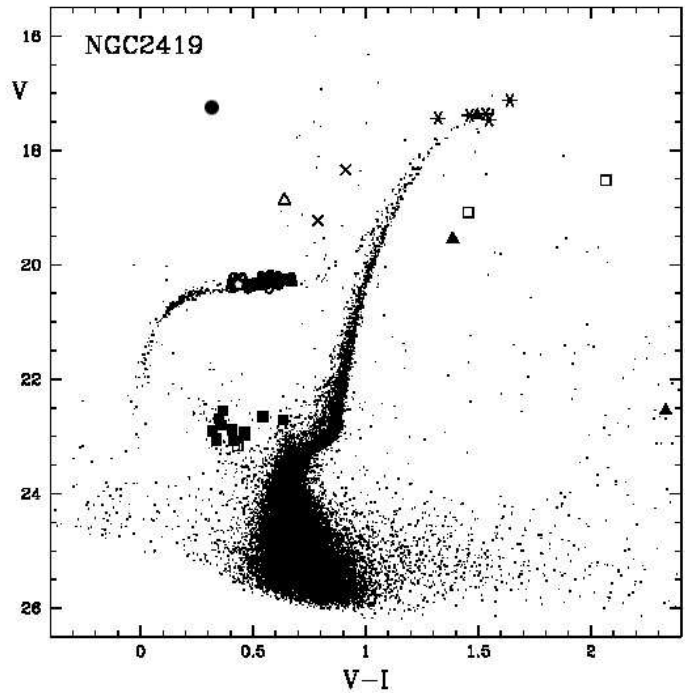}
\caption{$V$ vs $V-I$ CMD of NGC~2419 from the SUBARU dataset, with the cluster variables plotted according to their intensity-averaged
magnitudes and colors, and using different symbols for the various type of variables.
Filled circles: {\it ab-}type RR Lyrae stars (RRab); open circles: first overtone (RRc) RR Lyrae;  pentagon: double-mode (RRd) star; open triangle:
Population II Cepheid; filled squares: SX Phoenicis stars; open squares: binary systems; asterics: long period and semiregular variables;
crosses: $\delta$ Scuti stars;
filled triangles: variable stars with non reliable classification in type (NC).}
\label{cmd}
\end{center}
\end{figure*}
\begin{figure*}
\begin{center}
\includegraphics[width=15cm]{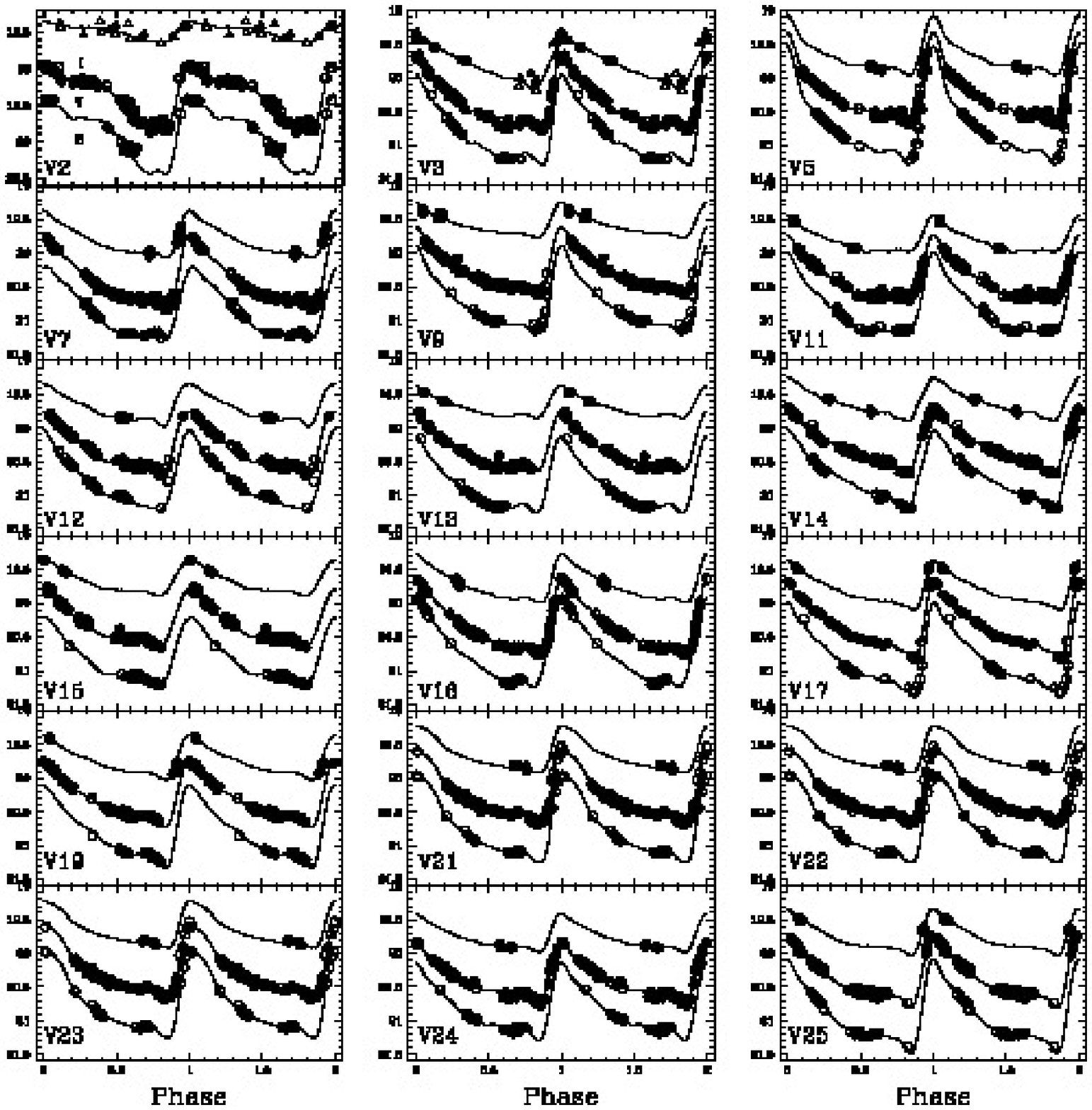}
\caption{$B,V,I$ light curves of fundamental-mode RR Lyrae stars in NGC~2419. Open and filled symbols are used for the TNG and SUBARU data, respectively.
Open triangles are the HST data. Lines are models obtained by properly scaling  the star's $V$ light curve according to the procedure described in
Section 3. The full catalogue of RRab light curves (38 stars) is published in the electronic edition of the paper. The errors on each phase point are of the same order of magnitude of the photometric accuracy reported in Table 1 for each telescope.}
\label{lc_RRab}
\end{center}
\end{figure*}
\begin{figure*}
\begin{center}
\includegraphics[width=15cm]{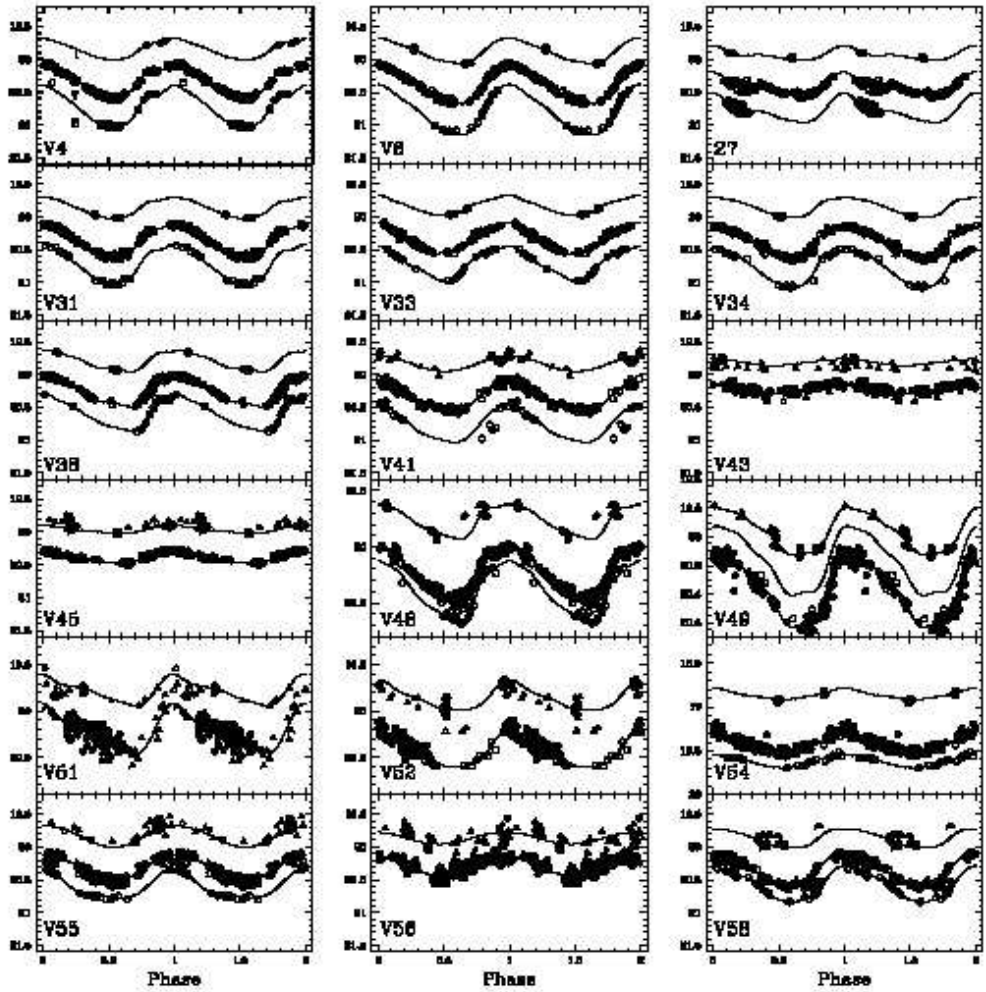}
\caption{Same as Fig. \ref{lc_RRab} but for first-overtone RR Lyrae stars. The full catalogue of RRc light curves (36 stars) is published in the electronic edition of the paper.}
\label{lc_RRc}
\end{center}
\end{figure*}
\begin{figure*}
\begin{center}
\includegraphics[width=15cm]{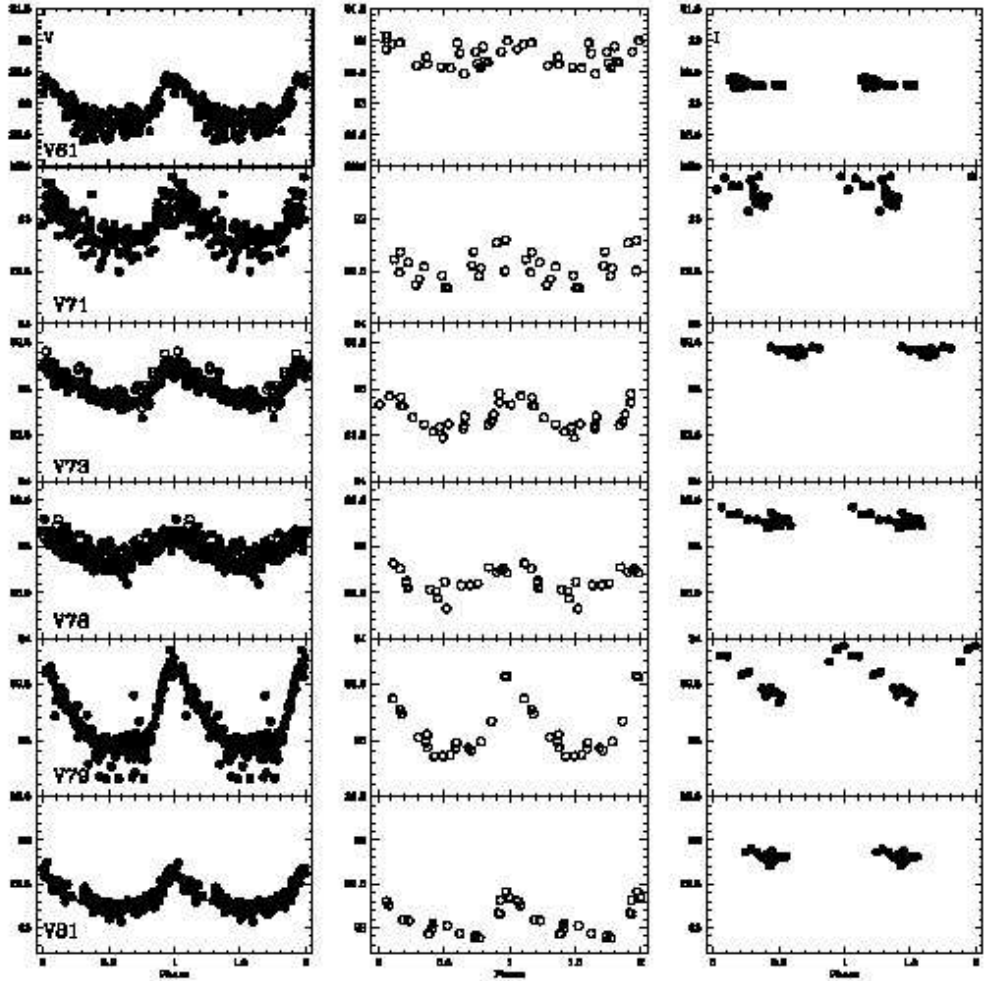}
\caption{$B,V,I$ light curves of SX Phoenicis stars. Symbols are as in Fig. 3. The full catalogue of SX Phe light curves (12 objects) is published in the electronic edition of the paper.}
\label{lc_SX}
\end{center}
\end{figure*}
\begin{figure*}
\begin{center}
\includegraphics[width=15cm]{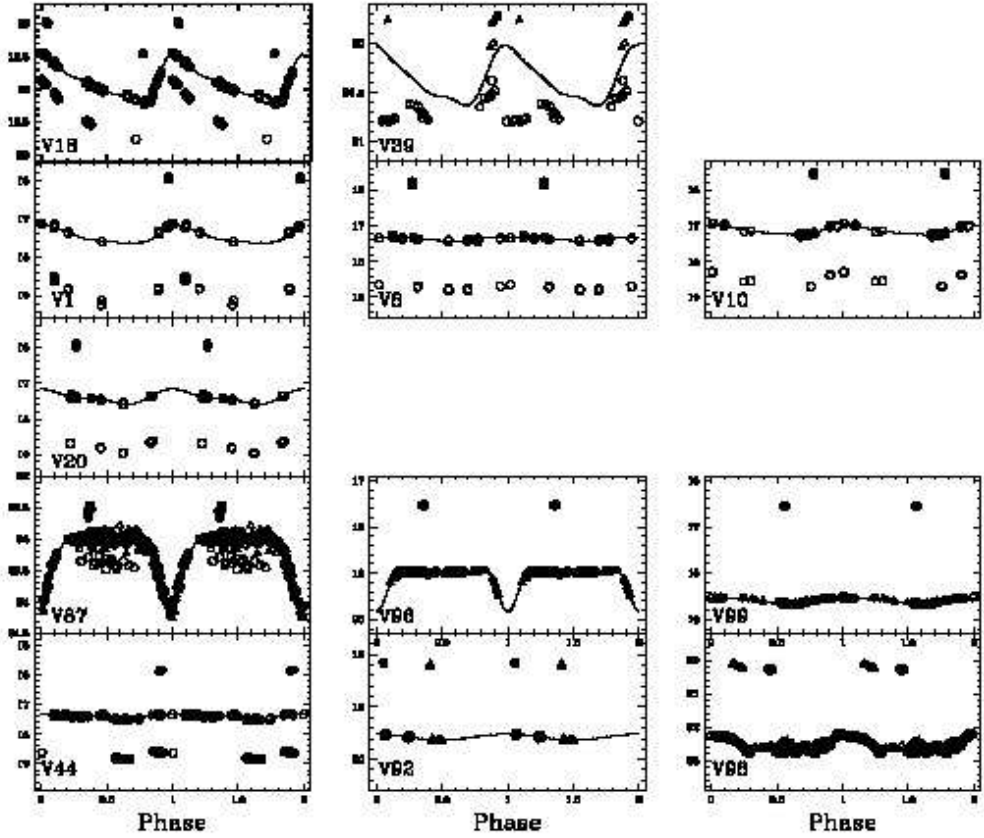}
\caption{
$B,V,I$ light curves of Pop. II Cepheids (V18), RRd (V39), LPVs (V1, V8, V10, V20 and V86), binary systems (V87, V96 and V99), non classified variables (V44, V92 and V98), and 
$\delta$Scuti stars (V95 and V100).
Symbols are as in Fig. 3.}
\label{lc_others}
\end{center}
\end{figure*}

Candidate variable stars were first identified on the $V$ frames of
NGC~2419 obtained with the TNG by using the Optimal Image Subtraction
technique as performed within the package ISIS 2.1
\citep{alard1,alard2}. %\citep{alard00}.
We identified about 300 candidate variables over the field of view
covered by the TNG observations, a large fraction of them were found
towards the cluster centre.  To increase the number of phase points
and obtain a better definition of periods and light curves, the TNG
candidate variables where then counteridentified on the HST and SUBARU
catalogs. Moreover, since the SUBARU observations better resolve the
cluster inner regions and, at the same time, cover a larger external
area compared to the TNG, an independent search for variables was carried out on the SUBARU $V$-band dataset by applying the following
procedure: i) we calculated the Fourier transform (in the
\citealt{sc96} formulation) of each star of the SUBARU photometric
catalogue for which we had at least 25 epochs; ii) we averaged this
transform to estimate the noise; 3) we calculated the signal-to-noise
ratio dividing by the noise the peak with maximum amplitude in the
transform.  We then cheched all the stars with high S/N, that for the
RR Lyrae stars typically means S/N in the range from 6.5 to 100.  All
candidate variables identified by any of the two independent
techniques described above were then checked for variability using
GRaTiS (Graphyc Analizer of Time Series), a custom software
developed at the Bologna Observatory by P. Montegriffo (see
\citealt{df99,clementini00}), which allows a period search using both the Lomb
periodogram (\citealt{lomb76}, \citealt{sca82}) and the best fit of
the data with a truncated Fourier series (\citealt{ba63}).  We
confirmed the variability of 101 stars, and derived period and
classification in type for 98 of them.  Whenever applicable, the study
of the light curves was performed by combining
data from the three different photometric catalogs (TNG, HST, SUBARU).
To combine the photometry we selected a number of reference stars
common to the three datasets, and studied the light curves using the
differential photometry with respect to the comparison stars.  The
total number of phase points of the combined datasets ranges from 17
to 20, from 22 to 212, and from 2 to 65 in the $B$, $V$ and $I$ bands,
respectively, with optimal sampling of the $V$ light curves of the RR
Lyrae stars, acceptable coverage in $B$, and rather poor sampling in
$I$, since the $I$-band observations are rather closely spaced in
time.  Periods were first derived from the $V$-band data alone, given
the better sampling, then applied to the $B$ and $I$-band data and
iteratively improved.  The large time baseline (ten years) covered by
the combined dataset, allowed us to solve alias ambiguities which were
rather severe when considering only individual datasets. The merged
datasets also returned a better estimate of the period, epoch of
maximum light and amplitudes of the light variation of the confirmed
variable stars.  Precision of the final period determinations is of
4-5 decimal places for variables with periods shorter than 2 days (95
objects) and increases up to 6 digits
for stars with the three datasets available.\\
Our sample of variable stars in NGC~2419 includes 76 RR Lyrae stars, 1
Population II Cepheid, 12 SX Phe stars, 2 $\delta$ Scuti variables, 3
binaries, 5 long period variables (LPVs) near the cluster RGB tip, and 3 stars which we were not able to classify in type. 
$V, B, I$ time-series photometry for each of the confirmed variables is
provided in Table 2 (the full table is available from the authors upon request). 
Fig.~2 shows the $V$ vs $V-I$ CMD of NGC~2419 obtained from the SUBARU dataset, with the variable stars identified in our study 
plotted according to their intensity-averaged
magnitudes and colors, and using different symbols for the various type of variables.\\
Identification and coordinates of the variable stars are provided in
Table \ref{tabellone}, along with type, period, time of maximum light,
intensity-averaged ${\rm \langle V\rangle}$, ${\rm \langle B\rangle}$
and ${\rm \langle I\rangle}$ magnitudes, and amplitudes of the light
variation (A$_V$, A$_B$ and A$_I$). For the double-mode star ($V39$), in the table we list the period corresponding to the first-overtone pulsation. 
Notes on individual stars are provided in Section 3.1.\\
Variables from V1 to V41 were already known from PR study, while stars
from V42 to V101 are new discoveries identified in the present
work. We assigned to the latter identification numbers increasing with
increasing the distance to the cluster center.  The complete atlas of
light curves is presented in Figs.~\ref{lc_RRab} and ~\ref{lc_RRc}
for fundamental-mode and first overtone RR Lyrae stars separetely, in
Fig.~\ref{lc_SX} for the SX Phe stars, and in Fig.~\ref{lc_others}, for the
other types of variables (namely Population II Cepheids, LPVs,  binaries, non classified 
and  $\delta$ Scuti variables).\\
The  good sampling of the $V$-band light curves allowed us to
obtain good best-fits of the visual data and a reliable
estimate of the average $V$ magnitude of the variable stars. On the other
hand, given the small number of $B$ and $I$ phase points and the
uneven distribution of the $I$-band observations, average $B$ and $I$
magnitudes are available only for a subsample of objects and are, in
general, more uncertain than the $V$ mean magnitudes.  To recover the
average magnitudes in $B$ and $I$ of the RR Lyrae stars, and thus be
able to plot them on the cluster CMD, when the $B$ and $I$ data
sampling was too sparse we used the star's $V$ band light curve as a
template and properly scaled it in amplitude to fit the $B$ and $I$
light curves. To constrain the scaling factors we computed
$A(B)/A(V)$, $A(V)/A(I)$ ratios using literature $B,V,I$ light curves of 130
RR Lyrae stars with good light curve parameters selected from the GCs
M68, NGC~5466, NGC~5053, M3, NGC~6362, NGC~6229, NGC~3201, M5,IC~4499
(see Clement catalogue available at http://www.astro.utoronto.ca/
$\sim$cclement/ for references on individual clusters)\footnote{The
  sample includes both {\it ab-} and {\it c-}type RR Lyrae variables
  and we verified that the derived ratio values are independent of
  type. The $A(B)/A(V)$ value is also in good agreement with the value
  defined for M15 \citep{corwin08}, whereas for $A(V)/A(I)$ we neglected a very small
  trend with amplitude present in some clusters for the RRab stars.}.
The derived ratios, computed as weighted averages, are $A(B)/A(V)$ =
1.29$\pm$0.02, and $A(V)/A(I)$ =1.58$\pm0.03$ independently of the
cluster metallicity and variable's pulsation mode.\\
Direct check of our scaling procedure and a fine tuning of the adopted
amplitude ratios was possible for the $B$ light curves because the
$\sim 20$ epochs available for this filter are evenly distributed and
allowed us to directly fit the observed $B$ light curves of the vast
majority of the RR Lyrae stars with very low errors ($<$ 0.01 mag)
adopting an $A(B)/A(V)$ ratio varying in the range from about 1.25 to
1.30.  No direct check was possible instead of the adopted $A(I)/A(V)$
ratio since the $I$-band light curves are, in general, too poorly
sampled, with all data points clustered around two separate
epochs. Thus, for this band we relied entirely upon the scaling value
derived from the literature data.  Additional checks showed that,
varying the $A(V)/A(I)$ ratio by 10\%, translates into an error of
about 3\% in $\left<I \right>$. Given the formal error of the
$A(V)/A(I)$ values ($\sim$2\%), we are thus confident that our procedure provides
accurate mean $I$ magnitudes, in spite of the uneven sampling of the
light curves in this band.  Indeed, once plotted on the cluster $V,
V-I$ CMD by using the $\left< V-I \right>$ colors derived with the
above procedure the RR Lyrae stars appear to fall very well into the
instability strip thus giving support to the reliability of our
procedure. The $A(V)/A(I)$ ratio derived for the RR Lyrae stars was also used to scale the light curves of the other types
of pulsating variables, since pulsation is caused basically by the same physical mechanism in all pulsating variables.
For the binary systems we adopted instead a ratio $A(V)/A(I)$ = 1
since variability in these stars
is due to a geometrical effect, hence is acromatic.\\
As a final note, due to the worse seeing conditions and the coarse
pixel scale of the TNG data with respect to the SUBARU images, the $B$
light curves of several RR Lyrae stars are contaminated by companions
and often have unreliable $\left< B-V \right>$ colors.  For this
reason we decided to plot the variable stars on the $V$, $V-I$ (SUBARU
dataset) CMD of NGC~2419 (Fig. \ref{cmd}).
%%%%%%%%%%%%
\begin{figure}[h]
\begin{center}
\includegraphics[width=8cm]{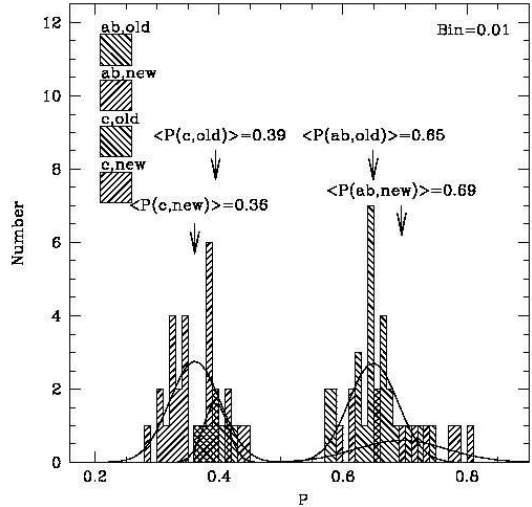}
\caption{Period distribution of {\emph ab-} and {\emph c-} type
\rrl\ stars, divided into known (ab-c, old) and newly discovered variables (ab-c, new). Periods are in days.}
\label{isto1}
\end{center}
\end{figure}
\subsection{Notes on individual stars}
{\bf V1, V8, V10, V20, V86}: these stars located around the cluster RGB tip are classified LPVs. They are saturated in all the HST frames and in the
SUBARU 180s long exposures. Only the TNG and SUBARU frames with $t_{exp}$=30s
where used to derive their light curves and pulsation properties.\\
{\bf V2}: this RRab is about one magnitude brighter than the HB, and being  at the center of the cluster it  is likely contaminated by companion stars which also may
cause the bumpy light curves from 0.2 to 0.6 in phase.\\
{\bf V27}: data from the TNG set are very scattered. The Fourier analysis gives several aliases which makes it difficult to  define the period. The star was also observed by \citet{baade}. The three sets of data (this paper, Baade, and PR) are not phased by the same period, probably because the period changed from P=0.35184 days by Baade,  to P=0.34896 days by PR, to the value P=0.34519 days inferred from our data.\\ 
{\bf V37}: there is a shift in magnitude between SUBARU and TNG light curves, we used both datasets to define the period, but only the SUBARU data to derive mean magnitudes and amplitudes.\\
{\bf V39}: double-mode RR Lyrae star with period ratio P1/P0= 0.745. 
In Table 3 we report the pulsation characteristics of the first-overtone (FO) pulsation, the pulsation characteristics of the fundamental mode are P$_{F}$=0.54621 d, A$_{V}$=0.347mag, A$_{B}$=0.454mag, A$_{I}=$0.305mag.\\
{\bf V44}: this star is very bright and is saturated in the HST frames and in the SUBARU 180 s exposures, only the TNG images and the SUBARU frames with $t_{exp}$=30 s could be used to derive its pulsation properties. The star has P$\sim$ 0.9 days and rather small amplitudes. According to the position on the CMD V44 falls in the region of 
LPVs.\\ 
{\bf V45}: the star is very close to the cluster center. The SUBARU data provide a $V$ amplitude larger than the TNG data. In Table \ref{tabellone} we have reported the visual amplitude inferred from the SUBARU data, the amplitude of the TNG data is $A_{V}> 0.416$mag.\\
{\bf V46}: the star is in the  central part of the cluster and likely contaminated by other stars. We have used only the SUBARU data because less scattered than the TNG observations.\\
{\bf V54}: the star, classified as first overtone RR Lyrae, has small amplitudes likely because contaminated by two companion stars. \\
{\bf V62}: The star is contaminated by a very luminous source. The SUBARU data are not usable and the TNG data are very scattered.\\
{\bf V70}: there is a shift in magnitude between TNG and SUBARU datasets for the star. In the present analysis we used only the TNG data which produce a smoother light curve.\\
{\bf V65, V71, V79, V76:} these stars are in the central part of the cluster.
There are systematic shifts among TNG, SUBARU and HST datasets of the four stars likely due to  crowding and blending effects. All three datasets were used to estimate the periods,
but only the SUBARU data to derive average magnitudes and amplitudes.\\
{\bf V80}: the star is near a very bright source. Only the TNG dataset was used since produced smoother light curves.\\
{\bf V83}: the star is in the blue straggler region and is tentatively classified as SX Phoenicis variable, however with its period of about 0.14 days
significantly deviates from the SX Phe PL relation (see Fig. 13).\\
{\bf V87}: binary system. In Table 3 we list the most probable period inferred for the star by the Lomb
periodogram since it gives less scatter than the period inferred from the Fourier analysis (P=0.184716 days, A$_{V}$=1.199 mag, A$_{B}$=0.166 mag, 
and A$_{V}$$>$0.265 mag).\\
{\bf V89}: the star has very small amplitudes for a c-type RR Lyrae. It lies very close to the edge of the frames and this may partially explain the anomalous
amplitudes.\\

\section[]{Properties of the NGC~2419 variable stars}

In their study of the NGC~2419 variable stars PR identified 32 RR
Lyrae stars, 1 Population II Cepheid, 4 red irregular/semiregular
variables, and further 4 variables on the HB, likely RR Lyrae stars,
for which they were not able to provide period and mode
classification. \citet{clement90} later found that V39, which PR classify as a first
overtone RR Lyrae star, is in fact %of
%first overtone RR Lyrae stars 
a {\it double-mode} pulsator
(RRd).  In our study, we recovered and derived reliable
periods for all the variables previously known  in NGC~2419 by PR study, and
detected 60 new variables that are mainly located in the cluster
central regions.
In the
following we discuss the properties of the various types of variables,
separately.
\begin{figure*}
%\vspace{-1.2cm}
\begin{center}
\includegraphics[width=11cm]{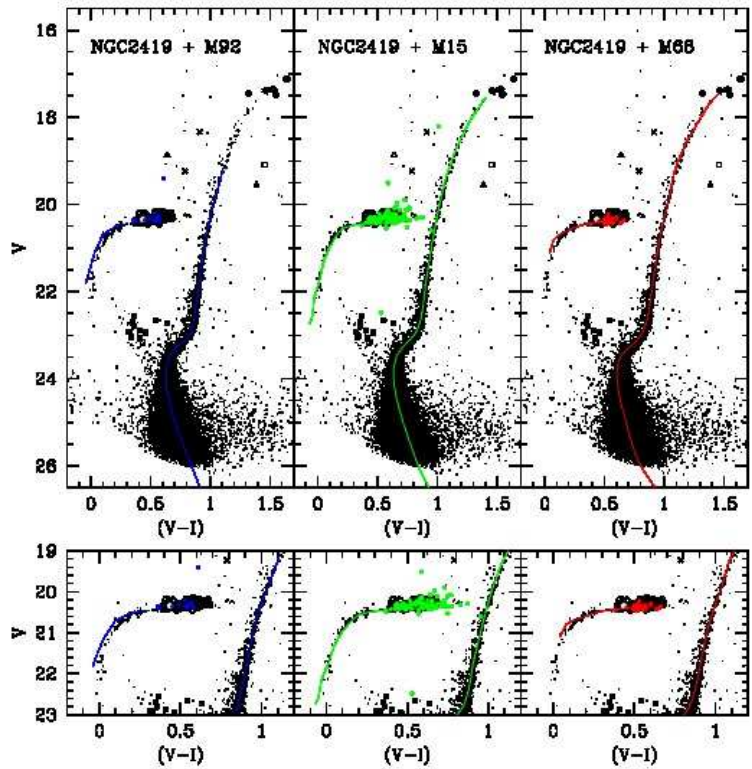}
%\vspace{-3.8cm}
\caption{Comparison of CMD, HB morphology and 
 distribution of the RR Lyrae stars on the HB of NGC~2419 with the mean ridge lines and RR Lyrae distribution of three 
 metal poor, OoII, Galactic GCs, namely, M92 (left panels), M15 (central panels), and M68 (right panels). Symbols are the same as in Fig. 2.
}
\label{figa6pannelli}
\end{center}
\end{figure*}
\subsection[]{The RR Lyrae stars}

PR classified as fundamental mode RR Lyrae stars (RRab) 25 of the
variables detected in NGC~2419 and further seven as first-overtone
pulsators (RRc). According to the average periods $\left< Pab
\right>$=0.654 days and $\left< Pc \right>$=0.384 days they concluded
that the cluster is an Oosterhoff type II (OoII) system (Oosterhoff 1939). 
This is consistent with the cluster low metal abundance
and HB morphology, and strenghtens the similarity with other three metal poor Galactic GCs, namely, M92, M15 and
M68, which are also OoII systems. We will come back to this point later in this section. 
 In our study we confirmed period and classification in types of the RR Lyrae stars found by PR.
Our analysis also confirmed both periods and period ratio:
P$_1$/P$_0$=0.745 found for the double-mode RR Lyrae star (V39) by
\citet{clement90}.  We also derived period and classification in
type for the 4 variables for which
PR do not provide an estimate of period, namely, V30, V38, V40, and V41.  We find that two are
$ab-$type and two are $c-$type RR Lyrae stars. As for the new
discoveries they include 11 fundamental mode and 28 first overtone
pulsators. The number of new {\it c-}type RR Lyrae stars discovered in
the present study is remarkably high, and increases by a factor 5 the
statistics of the RRc stars detected in NGC~2419 by PR.  On the other
hand, the new RRab stars extend the period range of the known
fundamental-mode pulsators in NGC2419 to longer periods, hence smaller
amplitudes, than in PR.  Most of the new RRc as well as the new RRab
stars with longer periods were missed by PR because their small
amplitudes and the location often towards the cluster center made them
rather difficult to detect with traditional techniques. % We discovered
Summaryzing, with the addition of the new
variables the number of RR Lyrae stars in NGC~2419 becomes: 38 RRab,
36 RRc and 1 RRd star, and the ratio of number of RRc over number of
RRc+RRab stars changes from 0.28 by PR to 0.49, in much better
agreement with expectations for an Oo II cluster.  The distributions in period
 of known and newly discovered RR Lyrae stars, are compared
in Fig. \ref{isto1} for {\it ab-} and {\it c-}type variables,
separately, and clearly show that PR missed the variable stars with smaller amplitudes.
The average period of the total sample of RR Lyrae stars is: $\left<
  Pab \right>$=0.662 d ($\sigma$=0.055, average on 38 stars), and
$\left<Pc\right>$=0.366 d, ($\sigma$=0.038, average on 36 stars), for
fundamental-mode and first-overtone pulsators, respectively.
The new values confirm and strenghten the classification of NGC~2419
as an OoII cluster.  The minimum period of the RRab stars,
$Pab_{min}$=0.576 d, is in perfect agreement with values of prototypes
OoII clusters like M15 \citep{corwin08} and M68 \citet{walker94}. In Fig. \ref{figa6pannelli} we compare the HB morphology and 
the distribution of the RR Lyrae stars on the NGC~2419 HB with the mean ridge lines and RR Lyrae distribution of three 
other metal poor, OoII, Galactic GCs, namely, M92 (left panels), M15
(central panels), and M68 (right panels of Fig. \ref{figa6pannelli}). 
The ridge lines for the three clusters were derived 
respectively from, \citet{johnson92,rosenberg00,walker}, while mean magnitudes
and colors for the RR Lyrae stars are from the studies of
\citet{kopacki,corwin08,walker}.
Ridge lines and RR Lyrae stars of M92 were shifted by $\Delta V$=5.25 mag and $\Delta (V-I)$=0.07 mag, 
those of M15 by $\Delta V$=4.55 mag and $\Delta (V-I)$=-0.005 mag, and those of M68 by $\Delta V$=4.75 mag and $\Delta (V-I)$=0.04 mag,
in order to match NGC~2419. 
This comparison confirms the similarity of  NGC~2419 to these three clusters. In particular, the ridge lines and HB morphology of M15 perfectly
matches the CMD and HB of NGC~2419. Moreover, the shift required to
match NGC~2419 allows us to get hints on the reddening affecting NGC~2419.
Indeed, adopting $E(B-V)=1.26 E(V-I)$, and the reddening reported by
\citet{harris96}: $E(B-V)_{M92}=0.02\pm0.01$ mag, 
$E(B-V)_{M15}=0.10\pm0.01$ mag, and $E(B-V)_{M68}=0.05\pm0.01$ mag, we
obtain $E(B-V)_{NGC~2419}=0.08\pm0.01$ mag. We note that this estimate is
intermediate between the value obtained by \citet{harris96}
($E(B-V)=0.11\pm0.01$ mag) from the average of a number of different
sources and the value derived
by \citet{schlegel} ($E(B-V)=0.065\pm0.010$ mag).\\
Finally, we note that the RR Lyrae instability strip (IS) of NGC~2419 is sharply defined and very
well sampled by the cluster variables.  Its boundaries are found at
${\rm \langle V\rangle - \langle I\rangle}$ = 0.40$\pm$ 0.01 mag (blue
edge), and ${\rm \langle V \rangle - \langle I\rangle}$ = 0.67$\pm$
0.01 mag (red edge). 

\subsubsection[]{The Bailey diagram}

\begin{figure}[h]
\vspace{-1.2cm}
\begin{center}
\includegraphics[width=8cm]{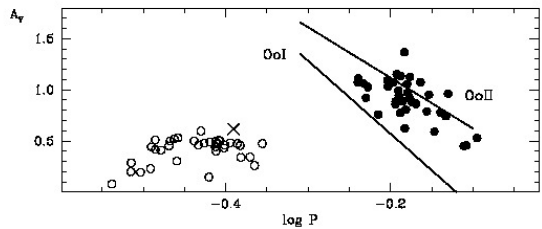}
\vspace{-3.8cm}
\caption{$V$-band period-amplitude diagram for RR Lyrae stars in NGC~2419. Filled and open circles
are {\emph ab-} and {\emph c-type} \rrl\ stars, respectively. The cross is the double-mode star.
The two lines are the positions of the OoI and OoII
GGCs according to \citet{clement01}. }
\label{figpa}
\end{center}
\end{figure}
The $V$-band period-amplitude distribution (Bailey diagram) of the
NGC~2419 RR Lyrae stars is shown in Fig.  \ref{figpa} and compared
with the loci of OoI and OoII GGCs taken from \citet{clement01}.  The
bulk of RRab stars clearly lies in the region defined by typical OoII
GGCs, confirming the classification of NGC~2419 as
OoII. The distribution of RRc stars shows a bell-shaped feature, as
expected for a low metallicity RR Lyrae population \citep{dc04}.  A few RRab
variables are scattered in the intermediate region between OoI and
OoII types.  The small amplitudes of some of these variables located
towards the cluster center may be due to contamination by companion
stars as in the case of V42 and V80 which both also have very
scattered light curves.
The other stars which do not show any particular spatial distribution 
could either represent the short-period tail of the OoII period
distribution, 
or could as well be undetected Blazkho variables (\citet{blazhko}).\\
In conclusion, according to our more complete census of the RR Lyrae
population, NGC~2419 confirms its nature of a {\em pure} OoII globular
cluster.
\subsection[]{The SX Phoenicis stars}
As recently pointed out by \citet{dalessandro} NGC 2419 contains one
of the largest BSS population ever observed in a globular cluster,
with more than 230 objects found in the brightest part of the
main-sequence.  Among them we detected 1 binary system, and 12
pulsating variables with periods in the range from 0.041 to 0.140d and
hints, in same cases, for the existence of several secondary
periodicities. They are likely SX Phoenicis stars.
Light curves for the SX Phe stars are shown in Fig. \ref{lc_SX}. Since
these stars obey a period luminosity (P-L) relation can
be used as distance indicator to estimate the cluster distance. This is presented in Section 6
where we also discuss the mode classification of these 12 variables.\\

\subsection[]{Others types of variable stars}

In Fig. \ref{lc_others} we show the light curves of other variable
stars in our sample, namely one Population II Cepheid (V18), the double pulsator (V39), 5 long period variables (LPVs) near the tip of the cluster RGB (V1, V8, V10, V20 and V86), 3 binary sistems (V87, V96 and V99). We also plot the light curves of   3 stars which we were not able to classify in type, as well as those of V95 and
 V100 which we classify as field $\delta$ Scuti stars, since 
   they have the typical period of this class of variables but are 
too bright to belong to NGC~2419.\\

\section[]{Distance to NGC~2419 from the cluster RR Lyrae stars}

In this section we calculate the distance to NGC~2419 using a number of methods based on the RR Lyrae stars.

\subsection[]{Mv-[Fe/H] relation}
The RR Lyrae stars are primary distance indicators for Population II systems, since their absolute magnitude
is almost constant and weakly dependent on metallicity. 
The average $V$ apparent  magnitude of the RR Lyrae stars can be used to derive
an accurate estimate of the distance to NGC~2419, provided that the cluster metallicity and reddening are known, and that a value
 is adopted for the absolute magnitude of the RR Lyrae stars. \\
According to our study NGC~2419 contains 75 RR Lyrae stars. Their mean
$V$ magnitude is ${\rm \langle V(RR)\rangle}$ = 20.31$\pm 0.01$ mag
($\sigma$=0.06, average on 67 stars) excluding 8 objects which are
either contaminated by companions or have incomplete light curves
(namely, stars V2, V70,V57, V54, V65, V80, V58, V32). This average value defines very
sharply the horizontal portion of the cluster HB.  The metal abundance
of NGC~2419 was estimated by \citet{suntzeff} to be [Fe/H]=$-$2.1
dex (on the Zinn \& West metallicity scale) using low resolution spectra. This value is in agreement with the metal abundance derived for the cluster by 
\citet{carretta2009c} in their newly defined metallicity scale.  Assuming for the
absolute luminosity of the RR Lyrae stars at [Fe/H]=$-1.5$,
$M_V$=0.54 mag, which is consistent with a distance modulus for the Large Magellanic Cloud 
 of 18.52 $\pm$0.09 mag \citep{clementini03}, 
$\Delta M_V/[Fe/H]$=0.214 ($\pm$ 0.047) mag/dex (Gratton et al. 2004) for the slope of the RR Lyrae luminosity
metallicity relation, $E(B-V)$=0.08$\pm$ 0.01 mag, and [Fe/H]=$-2.1$ dex, the
distance modulus of NGC~2419 derived from the mean luminosity of its
RR Lyrae stars is (m-M)$_0$=19.65 $\pm$ 0.09 mag  
(D= 85.1 $\pm$ 3.5 kpc). 
The
final errors on the distance include the standard deviation of the
mean HB magnitude and the uncertainties in the photometry, reddening,
and RR Lyrae absolute magnitude. 
The present value is in good agreement with literature values \citep{harris96} for the distance to NGC~2419 and is about 0.05 mag longer than 
derived by Ripepi et al. (2007), due to the adoption of a different value for the absolute magnitude of the RR Lyrae stars.
\\
\subsection[]{Pulsation distances}

\begin{figure}[h]
\begin{center}
\includegraphics[width=8.2cm]{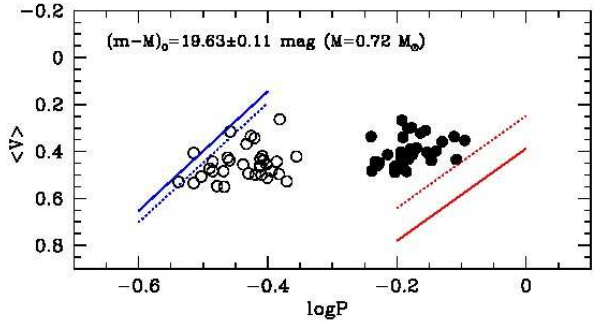}
\vspace{-3.6cm}
\caption{RR Lyrae stars of NGC 2419 in the $M_V$ vs $\log P$ plane. Filled and open circles
are RRab and RRc pulsators, respectively. Solid  and dashed lines
are the theoretical boundaries of the instability strip calculated
for values of the mixing length parameter of $l/H_p$=1.5 and $l/H_p$=2.0, respectively.}
\label{FOBE}
\end{center}
\end{figure}
\begin{figure}[h]
\vspace{-3cm}
\begin{center}
\includegraphics[width=8cm]{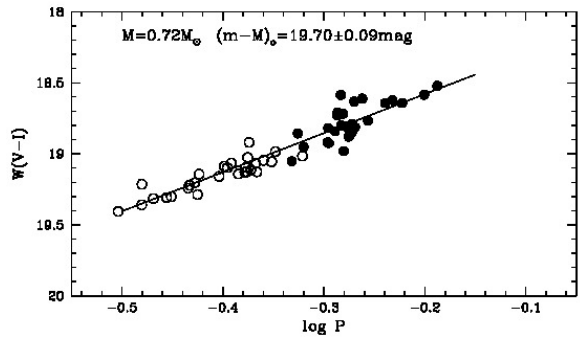}
\caption{Comparison between observed and predicted Period-Wesenheit relation (symbols are as in Fig. \ref{FOBE}). Periods of c-type pulsators were fundamentalized 
by adding the value +0.127 to the logarithm of the star period (Di Criscienzo et al. 2004).
The solid line represents the predicted fundamental-mode relation for M=0.72 M$_{\odot}$}
\label{WES}
\end{center}
\end{figure}

\begin{figure}[h]
\vspace{-3cm}
\begin{center}
\includegraphics[width=8cm]{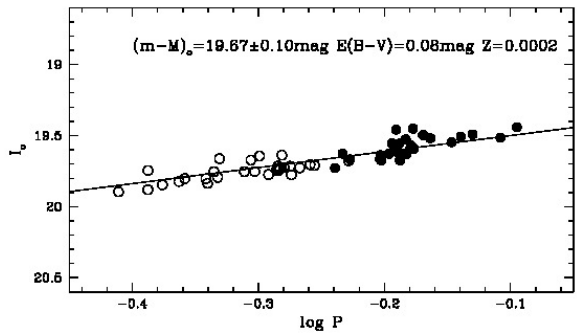}
\caption{Comparison between predicted and observed RR Lyrae period-luminosity relation in the $I$ band (symbols are as in Fig. \ref{FOBE}). Periods of the c-type
variables were fundamentalized by adding the value +0.127 to the logarithm of the star period (see Di Criscienzo et al. 2004). The solid line shows the predicted relation for Z=0.0002.}
\label{pli}
\end{center}
\end{figure}

We have estimated the distance to NGC~2419 with further four
methods: i) the comparison of theoretical and empirical RR
Lyrae first overtone blue edges (FOBE) and fundamental red edges (FRE)
in the $M_V$ versus $\log P$ plane \citep{caputo97,caputo00,dc04}; ii) the theoretical Wesenheit function \citep{dc04}; 
iii) the Period-Magnitude in the $I$ band by \citet{catelan04b}; and iv) the light curve model fitting method \citep{bono2000}. 
%As an additional constraint to the cluster distance modulus, at the end of this section we will also present results from the model fitting of the observed light curves of three RR Lyrae stars.\\
In order to compare the observed
properties of RR Lyrae stars with the model predictions, we need an
estimate of the mean mass of the pulsators. To this aim we can use the
results of synthetic HB models as a function of HB type and global
metallicity Z (see \citealt{dc04}), hence, we have to convert the observed iron abundance ([Fe/H]=-2.1 dex)
into global metallicity Z. This trasformation can be tricky due to
the uncertainty of the $\alpha$ enhancement for NGC~2419, for
which there are no direct measurements available. Taking into account the total
range of $\alpha$ enhancement spanned by GGCs: $\alpha$=0 and $\alpha$=0.5
\citep{gratton04}, we obtain a range of Z=0.00015$\div$0.00035 for the total metal abundance. 
We have then re-evaluated the cluster HB type (HB$_{Lee-Zinn}$) based on our CMD of NGC~2419.
 This is the so-called Lee-Zinn parameter
  used to classify the morphology of the HB and depends on the
  number of stars inside and on either sides of the RR Lyrae IS
  (HB$_{Lee-Zinn}$=(B - R)/(B + V + R)). We have derived its value using the SUBARU data and considering only RR Lyrae stars outside an area of 50 arcsec in radius from the
cluster center (40 RR Lyrae stars in total). We took off the cluster center to alleviate problems related to blending and incompleteness.  
We obtain an HB type equal to 0.80 if we exclude the blue hook stars from the number B of stars bluer than the 
blue edge of the IS,
0.83 if we include also the blue hook stars, and 0.75 if only consider blue stars with
V brighter than $V$=21 mag. In the end we chose to use the value 0.80 for the HB type. It should be noted  that at 
the metal abundance of NGC~2419 the HB type does not affect significantly the choice of the value for the 
mass \citep[see Fig. 8 of][]{dc04}.
Using the above values for the global metal abundance, and our estimate of the cluster HB type we derive that
the average mass of NGC~2419 RR Lyrae stars is in the range from 0.74 to
0.69 M$_{\odot}$. Using instead the period ratio of the double-mode RR Lyrae star P$_1$/P$_0$=0.745 and the Petersen diagram as suggested by \citet{bragaglia2001} and \citet{bono96} we would derive a larger value of the mean mass of 
$\sim$0.80 M$_{\odot}$.
To take in account the  uncertainty in the mass using a conservative approch we adopted a mass of M=0.72 M$_{\odot}$ and varied this value by
$\pm$0.05 M$_{\odot}$ when
calculating quantities depending on mass.\\

\subsubsection{FOBE and FRE}
To apply this method we have first calculated the theoretical
absolute visual magnitude of first-overtone blue edge and 
fundamental red edge, namely M$_V$(FOBE) and M$_V$(FRE), by means of equations 2 and 3 in \citet{dc04}. 
 These relations depend on the
stellar mass and mixing lenght parameter $l/H_p$.  For the mass value
we used 0.72 M$_{\odot}$, as discussed above, while for $l/H_p$ we
followed the prescriptions by \citet{dc04}, i.e $l/H_p$=1.5 for the FOBE
and $l/H_p$=2.0 for the FRE. The results of this procedure are shown in
Fig. \ref{FOBE}.  The distance modulus required to match the observed
boundaries of the IS with the theoretical relations is
(m-M)$_0$=19.63$\pm$0.11 mag, where the error takes into account a
variation of 0.05$M_{\odot}$ in stellar mass, an error of 0.04 in
the mixing lengh parameter, and the standard
deviation of the relations (0.07 dex).\\

\subsubsection{Wesenheit function}
This second method uses the predicted Wesenheit function in the $V$ and
$I$ filters\footnote{We have chosen to use the $V$ and $I$ bands because the SUBARU data have  higher
  quality  compared to the TNG $B$ data.}. According to
equation 4b in \citet{dc04}, to apply this relation we need three
observed quantities, namely the mean $V$ magnitude, the mean $(V-I)$ color, and the period of each RR
Lyrae star (once the mass is fixed at 0.72 M$_{\odot}$). The
application of the method to the NGC~2419 RR Lyrae stars is shown in
Fig. \ref{WES} where the c-type RR Lyrae stars were fundamentalized by adding +0.127 to the logarithm of the first overtone period (see Di Criscienzo et al. 2004). The resulting distance modulus is
(m-M)$_0$=19.70$\pm$0.09 mag, where again the error takes into account
the error in mass, and the intrinsic dispersion of the
adopted theoretical relation ($\sim$0.03 mag).\\

\subsubsection{Period-Luminosity relation}
The third method involves the use of the Period-Luminosity (PL) relation in the $I$
band. This correlation is not as tight as the $K$ band (PL), but is sufficiently
well defined to be used as a reliable distance indicator. In this work we used
the relationship given by Catelan et al. (2004), based on synthetic HB models,
which essentially depends on the overall metallicity Z. 
The derived distance modulus adopting Z=0.0002 is {\rm (m-M)$_0$ =19.67 $\pm$ 0.10 mag where the error is mainly due to uncertainty on the adopted Z. We note that the value adopted for Z falls in the range derived in Section 5.2.}
%The result of this fit
%is shown in Fig. \ref{pli}. The derived distance modulus adopting Z=0.0002\footnote{This value was  chosen beacuse whithin the observed measures is that which falls within the range derived in Section 5.2.}
%is (m-M)$_0$=19.67$\pm$0.10 mag where the error is mainly due to the uncertanty on the adopted Z.\\

\subsubsection{Ligh curve model fitting}
Further constraints on the distance to NGC~2419 can be obtained
  by the model fitting of the observed light curves with nonlinear convective hydrodynamical models (see e.g. Di Criscienzo et al.
2004; Bono, Marconi \& Stellingwerf 1999).
This very promising technique was already applied with success to field
and cluster RR Lyrae stars by \citet{bono2000,MarClem,MarScilla}.
In the case of the RR Lyrae stars in NGC~2419  we have well sampled light curves in at
least  one photometric band (the $V$ band), allowing in principle to apply the method to
an extensive and homogeneous sample of pulsators. For a preliminary application
in this paper we have selected one fundamental mode RR Lyrae star, variable V3,  and two first overtone RR Lyrae stars, V34 and V75.

\begin{figure}[h]
\begin{center}
\includegraphics[width=8cm]{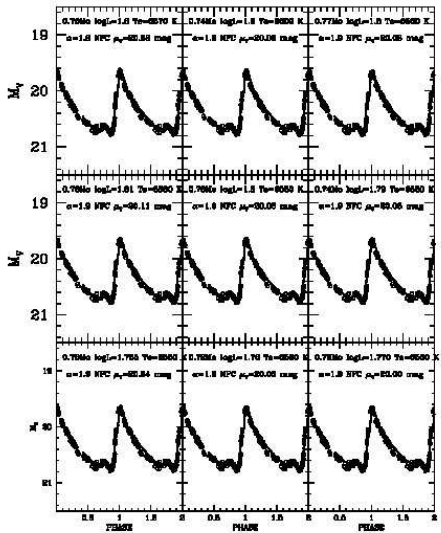}
\caption{Results from the theoretical modeling of the observed $V$-band
  light curve for the RRab star V3. The best fit model parameters: effective temperature, mass and luminosity are labelled together with the 
adopted mixing length value and the inferred $V$-band apparent distance
  moduli, assuming $E(B-V)$=0.08 mag.
}
\label{V3-V75}
\end{center}
\end{figure}

\begin{figure}[h]
\begin{center}
\includegraphics[width=8cm]{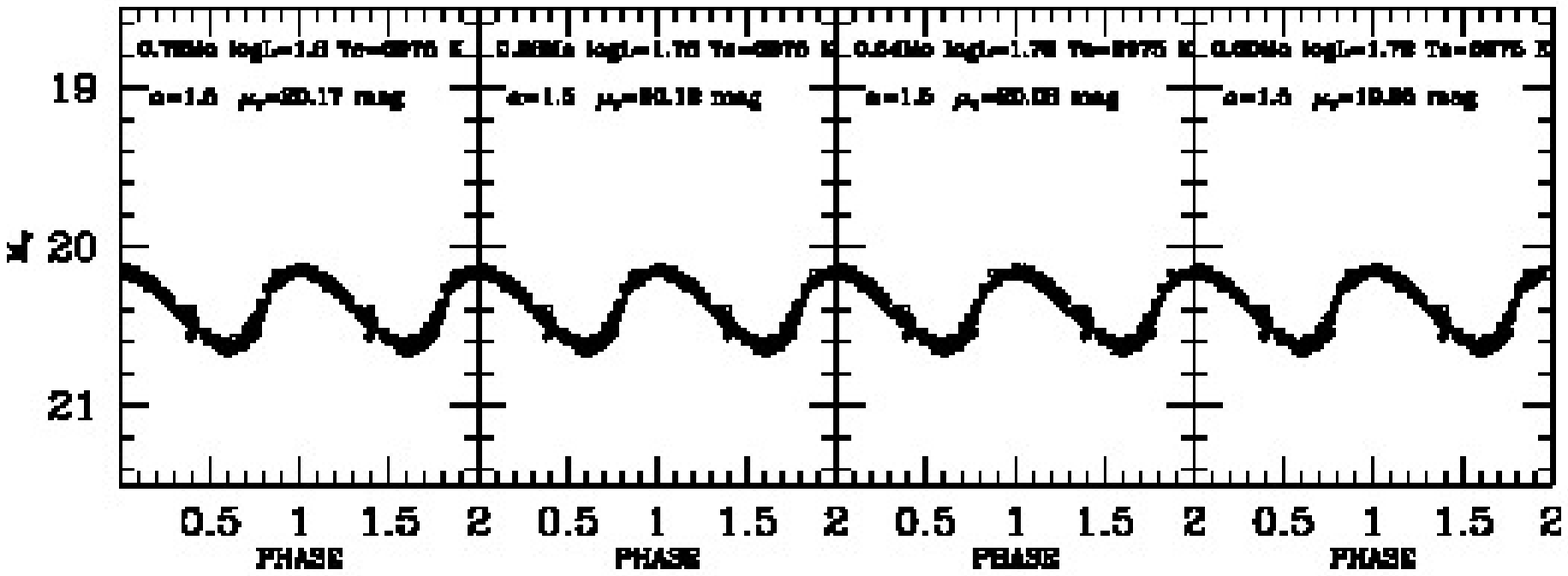}
\vspace{-5cm}
%\includegraphics[width=15cm]{../FIGURE_AJ/V20587new.ps}
%\vspace{-9cm}
\caption{Same as figure Fig. \ref{V3-V75} for the RRc star V34.}
\label{V34}
\end{center}
\end{figure}
\begin{figure}[h]
\begin{center}
\includegraphics[width=8cm]{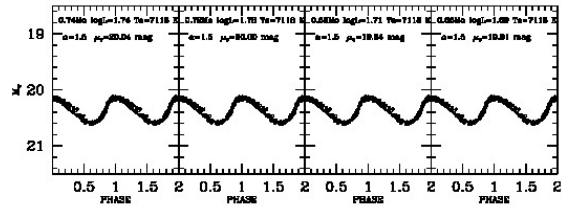}
\vspace{-5cm}
\caption{Same as figure Fig. \ref{V3-V75} for the RRc star V75.}
\label{V75}
\end{center}
\end{figure}

By using the procedure outlined in \citet{MarClem,MarScilla}
 and references therein, for 
the three 
pulsators we computed a set of pulsation models varying the
stellar parameters in order to reproduce the period, the amplitude and
the morphology of the observed light curves. In particular, we
fixed the chemical composition (Z=0.0002, Y=0.24\footnote{We chose this value for the helium abundance as it is close to the Big Bang value (Coc et al. 2004).}) and allowed the mass
(and, as a consequence, the luminosity at fixed effective temperature)
to vary in the interval suggested by synthetic HB models (see the discussion
on the mass given at the beginning of this section).  As a result we defined
the best effective temperature for which the 
pulsation models are able to reproduce the observed light variations
within $\pm$ 0.05 mag over all the pulsation cycle for stellar masses
and luminosities consistent with the evolutionary prescriptions.  In
Figs. \ref{V3-V75},  \ref{V34}, and \ref{V75} we show the best fit solutions (solid lines)
overimposed to the observed $V$ band light curves. The inferred stellar
parameters and distance moduli are labelled in each panel. 
In the case of V75 we find, É. of $\pm$0.07 mag on the inferred apparent distance modulus of 19.73 mag. For V34 we find, É. of $\pm$ 0.11 mag on the inferred apparent distance modulus of 19.83 mag, whereas
for the RRab star V3 an uncertainty of $\pm$ 0.04 in mass causes causes an error of +/-0.08 mag on the derived distance modulus of 19.81 mag.
The errors on the inferred
distance moduli have to be increased to take into account also the
contribution of photometric and reddening uncertainties.  \\
\subsubsection{Summary of the distance estimates}
Distances to NGC~2419 obtained by the various methods described above are summarized in Table \ref{moduli}:  all these estimates are consistent
to each other within their uncertainties. Although results from the model fitting tend to provide fainter distance moduli than the other methods.
A mean of the values in Table  \ref{moduli} weighting each individual value by its error leads
the mean distance modulus of $\mu_0$ (NGC~2419) = 19.714 $\pm$ 0.078 mag, while a simple mean would lead to 
$\mu_0$ (NGC~2419) = 19.717 $\pm$ 0.078 mag. A mean computed discarding the longest and shortest moduli would lead 
to  $\mu_0$ (NGC~2419) = 19.706 $\pm$ 0.063 mag.
Thus, our final adopted distance modulus for NGC~2419 from the RR
Lyrae stars is $\mu_0$ (NGC~2419) = 19.71 $\pm$ 0.08 mag (D= 87.5
$\pm$ 3.3 kpc) which agrees within the errors with the value found by Harris et al. 1997 ($\sim$90 kpc).

\section[]{The SX Phoenicis distance to NGC~2419}

\begin{figure}[h]
\vspace{-4cm}
\begin{center}
\includegraphics[width=8cm]{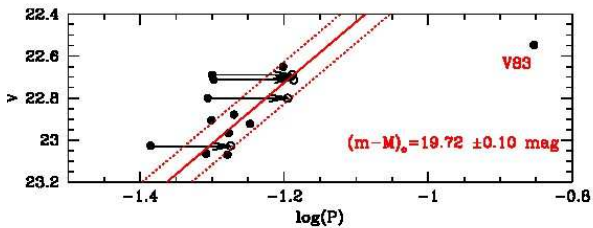}
\vspace{-2.5cm}
\caption{Position of the SX Phe stars in the $V-\log P$ plane (filled
  circles). Open circles are variables ``fundametalized'' according to
  the procedure described in the text. The solid line is the PL
  relation by McNamara et al. (2004) using the labelled value of
  distance modulus.}
\label{sx1}
\end{center}
\end{figure}
The mean apparent $V$ magnitudes of the 12 SX Phoenicis stars discovered in NGC~2419 are
plotted versus the observed periods (filled circle) in
Fig. \ref{sx1}. In the figure open circles represent variable stars 
``fundametalized'' according to the method described by various
authors in the literature (see e.g. McNamara, Clementini \& Marconi
2007). Following these authors stars brighter than the PL relation in the
period-magnitude plane are pulsating in a different  pulsation mode. In
particular, by adopting the slope of the PL relation by
\citet{mcnamara04} the SX Phoenicis stars lying above the relation in the
period-magnitude plane (Fig. 16) are first overtone (FO) pulsators. As
P$_{FO}$/P$_{F}$=0.775 (Poretti et al. 2005), these stars must be 
shifted rightwards by 0.111 dex in order to 
fundamentalize them (open circle in Fig. \ref{sx1}). Applying this
shift all the stars seem to lie on the same well defined PL
relation. Using the PL slope by McNamara et al. (2004), and E(B-V)=0.08 we
obtain $\mu_0$=19.72 $\pm$ 0.10 mag (see also Fig. \ref{sx1}) in good
agreement with the distance values found with methods based on the RR Lyrae stars.

\section{Conclusions}

We have presented a new photometric study of the variable stars in the remote Galactic
globular cluster NGC 2419, using proprietary and archive $B,V,I$
time-series CCD photometry of the cluster which span a large time interval and cover a total area 
of 50 $\times $43 arcmin$^{2}$ centered on NGC~2419 (see Paper I).
Variable stars were identified using the image subtraction technique.  
We recovered and confirmed periods for all the variables previously known in the cluster (41 objects) by PR photographic study, 
as well as discovered 60 new variables, doubling the number of RR Lyrae
stars. The new discoveries include 11 fundamental mode and 28
first overtone pulsators increases by a factor 5 the statistics of the
RRc stars detected in NGC~2419 by PR. Most of the new
RRc stars as well as the new RRab stars with longer periods were missed by
PR because their small amplitudes and the location often towards the
cluster center made them rather difficult to detect with traditional techniques and photographic data.
In our study we also detected 12 cluster SX Phoenicis stars. 
In this paper we have presented the catalogue of light light curves for all the 101 variables of NGC~2419. 
and studied in detail their properties.
Both RR Lyrae and SX Phe stars were used to estimate the cluster distance by means of a variety of different and in same cases indipendent methods.
The final distance modulus derived for NGC~2419 from its  RR Lyrae population is:  $\mu_0$ (NGC~2419) = 19.71 $\pm$ 0.08 mag  corresponding to D= 87.5 $\pm$ 3.3 kpc, in good agreement with the modulus derived from the cluster SX Phe stars and results from other methods in the literature. 
 The improved number of RR Lyrae stars and a
  detailed analysis of their pulsation properties were used to re-evaluate the Oosterhoff classification of NGC~2419 and get
  hints on the intriguing nature of this cluster. 
We confirm a ``pure"  OoII classification for NGC~2419 as also suggested by thecluster metallicity, this would disfavor an extragalactic origin for NGC~2419, since  RR Lyrae stars in extragalactic systems (e.g. field and cluster RR
Lyrae stars in the Magellanic Clouds and in classical dwarf
spheroidal galaxies)  generally have properties intermediate between
OoI and II types \citep{catelan04,catelan05}. 
However, the OoII behaviour of NGC~2419, by itself, is
not a sufficient condition to conclude that the origin of NGC
2419 is not extragalactic. Indeed, 5 of the LMC GCs and the Ursa
Minor dSph also have OoII type. Moreover, most of the ultrafaint dwarf
spheroidal galaxies recently discovered by the SDSS (e.g. Bootes I, Canes Venatici~II, Coma,
Leo IV, and UMAII) are found to be pure Oosterhoff type II systems
\citep{cseres01,dallora06,greco08,musella08,moretti09}.\\
In conclusion current results for NGC 2419 stellar content
(including the variable stars) do not allow us to explain why this
cluster is anomalous both as a globlular  cluster and as a possible
remnant of a distrupted dwarf galaxy, pointing toward the need of a
detailed analysis of the CMD and in particular of its HB.\\
Our data providing  a first complete mapping of the cluster HB, from the
extreme ``blue hook'' to the red portion, have opened the possiblity for 
such a study through theoretical evolutionary models and synthetic populations. Such a study is under way and will be
of great importance to put constraints of the origin of NGC~2419.

\acknowledgments Financial support for this study was provided by
  PRIN MIUR 2007 (P.I: Piotto). M.D.C thanks the Osservatorio di
  Bologna for hospitality during her stay when part of this work
  was done and  F. D'Antona for very interesting discussions on the still mysterious nature of NGC 2419.
\begin{table*}[h]
\fontsize{6}{6}
\caption{Instrumental set-ups and logs of the observations.}
\begin{tabular}{lllllrrrc}
\hline
\hline
\noalign{\smallskip}
{\rm Dates} & {\rm Telescope} & {\rm ~~Instrument} &{\rm ~Resolution}            & {\rm ~~~~FOV}&{\rm $N_ B$} &{\rm $N_ V$}& {\rm $N_ I$}&{\rm Photometric~accuracy}\\
            &                 &                    &                             &              &           &          &           &{\rm (HB~level)}          \\
{\rm UT}    &                 &                    &{\rm ~~($\prime \prime/pixel$)}&              &           &          &           & {\rm (mag)}              \\ 
\noalign{\smallskip}
\hline
\noalign{\smallskip}
{\rm Sept.,2003 - Feb., 2004}& {\rm TNG~~~} & {\rm  ~~~Dolores }   &~~~0.275&9.4$^{\prime}$ $\times$ 9.4$^{\prime}$     & 20~   &22~ &\nodata& 0.01-0.03\\
{\rm May, 1994 - Mar., 2000} & {\rm HST~~~} & {\rm  ~~~WFPC2   }   &~~~0.1  &$\sim$ 2.5$^{\prime}$ $\times$ 2.5$^{\prime}$& \nodata&18~ &10& 0.1-0.15\\
{\rm Nov., 1997}             & {\rm HST~~~} & {\rm  ~~~WFPC2   }   &~~~0.1  &$\sim$ 2.5$^{\prime}$ $\times$ 2.5$^{\prime}$& \nodata&7~~ &39& 0.1-0.15\\
{\rm Dec., 2002}             & {\rm SUBARU} & {\rm  Suprime-Cam}   &~~~0.20 &~34$^{\prime}$ $\times$ 27$^{\prime}$        & \nodata&165~&16& 0.01-0.02 \\
\hline
\label{tobs}
\end{tabular}
\end{table*}

\begin{table*}
\begin{center}
\caption{$V,B,I$ time series photometry of the variable stars. A portion of Table 2 is shown here for guidance regarding its form and content. The entire catalog is available from the authors upon request.}
%in the electronic edition of the journal.}
\begin{tabular}{cccccc}%\setlength{\tabcolsep}{3pt}
\hline
\hline
\multicolumn{4}{c}{Star V3 - {\rm RRab}} \\
\hline
{\rm HJD} & V  & {\rm HJD } & B & {\rm HJD } &  I \\
{\rm ($-$2452609)} &   & {\rm ($-$2452905) } & &{\rm ($-$2452609) } &  \\
 {\rm (days)}& {\rm (mag)}  & {\rm (days) } & {\rm (mag)}  & {\rm (days) } &  {\rm (mag)} \\
\hline
 0.051979  &  20.303    &    0.746271  &  20.256    &   0.038409  &  20.124  \\
 0.053994  &  20.294    &    1.754012  &  21.198    &   0.039415  &  20.009  \\
 0.055001  &  20.450    &   86.592255  &  20.666    &   0.040420  &  20.040  \\
 0.056019  &  20.487    &   86.606125  &  20.714    &   0.041426  &  19.929  \\
 0.057028  &  20.034    &   86.619461  &  20.796    &   0.042435  &  20.107  \\
 0.058034  &  20.070    &   86.628792  &  20.816    &   0.043442  &  20.030  \\
 0.059041  &  20.046    &   86.638573  &  20.835    &   0.044447  &  20.190  \\
 0.061506  &  20.068    &  127.485786  &  21.189    &   0.045453  &  19.502  \\
%
% 0.062513  &  20.076    &  127.498947  &  21.195    &   0.890455  &  19.442  \\
% 0.063521  &  20.052	 &  127.511253  &  21.172    &   0.896011  &  19.446  \\
% 0.064528  &  20.064	 &  127.520966  &  21.205    &   0.934208  &  19.327  \\
% 0.065534  &  20.070	 &  127.530174  &  21.191    &   0.953654  &  19.383  \\
% 0.066542  &  20.117	 &  127.539642  &  21.212    &   0.959210  &  19.446  \\
% 0.067548  &  20.086	 &  148.543991  &  20.584    &   0.960599  &  19.379  \\
% 0.068555  &  20.066	 &  148.564499  &  20.688    &   1.030165  &  19.392  \\
% 0.070325  &  20.111	 &  148.575409  &  20.707    &   1.072529  &  19.378  \\
% 0.073810  &  20.102	 &  148.584717  &  20.750    &   1.082252  &  19.415  \\
% 0.076508  &  20.098	 &  148.595490  &  20.801    &   1.087808  &  19.315  \\
% ...       &  ...       &  ...         &  ...       &   ...       &  ... \\
\hline
\end{tabular}
\end{center}
\label{hjd}
\end{table*}
\begin{deluxetable}{ll|cc|lll|cccccc|}
\tablecolumns{13}
\setlength{\tabcolsep}{0.04in} 
\tablewidth{0in} 
\tabletypesize{\scriptsize}{}
\tablecaption{Identification and properties of NGC 2419 variable stars. Notes on individual stars are provided in Section 3.1.}
\tablehead{Name   &    Id    &  RA(J2000) & DEC(J2000) & Type   &  P     &  Epoch      &  $\langle V\rangle$     &    $A_V$     &  $\langle B\rangle$   &  $A_B$    &   $\langle I\rangle$ & $A_I$}
\startdata
 V01   &    17997    &    7:38:11.611  & 38:51:58.95 &  LPV   &  198.0     &  52617.00      &    17.437   &    $\geq$0.477   &     18.925   &  $\geq$0.691    &   16.116    & indef \\
 V02   &    23327    &    7:38:07.955  & 38:52:33.67 & RRab      &  0.670170  & 53030.5599     &    19.331   &    0.881     &     19.845   &  $\geq$0.781   &   18.481    & $\geq$0.232 \\
 V03   &    22280    &    7:38:12.705  & 38:52:27.36 & RRab      &  0.625986  & 52617.0987     &    20.344   &    1.091     &     20.755   &  $\geq$1.101    &   19.783    & $\geq$0.781  \\
 V04   &    23625    &    7:38:15.128  & 38:52:35.12 & RRc       &  0.39210   & 52609.1238     &    20.317   &    0.502     &     20.698   &  0.672    &   19.843    & indef \\
 V05   &    33749    &    7:38:11.142  & 38:53:38.78 & RRab      &  0.655868  & 52608.667      &    20.180   &  $\geq$1.048 &     20.591    &  $\geq$1.334   &   19.645    & indef \\
 V06   &    10150    &    7:38:12.889  & 38:50:43.78 & RRc       &  0.371609  & 52617.1148     &    20.373   &    0.596     &     20.749   &  $\geq$0.638   &   19.871    & $\geq$0.276   \\
 V07   &    39159    &    7:38:16.185  & 38:54:18.79 & RRab      &  0.627331  & 52611.095      &    20.357   &    1.088     &     20.822   &  $\geq$0.766   &   19.793    & $\geq$0.267  \\
 V08   &    33054    &    7:38:06.842  & 38:53:34.14 & LPV   &   13.0     &   52581.470    &    17.362   &    0.118     &     18.722   &  0.141    &   15.830    & indef    \\
 V09   &    21821    &    7:38:05.633  & 38:54:21.16 & RRab      &   0.644733 &  52608.365     &    20.219   &    $\geq$0.798 &   20.671   &  $\geq$0.556    &   indef    & indef  \\
 V10   &    18271    &    7:38:09.890  & 38:52:01.07 & LPV   &  19.05     &  53053.400     &    17.123   &    0.344     &     18.542   &   $\geq$0.419   &   15.483    & indef  \\
 V11   &    24835    &    7:38:16.425  & 38:52:42.26 & RRab      &  0.58918   & 52609.0141     &    20.339   &  $\geq$0.779 &     20.753   &  $\geq$0.342    &   19.809    & $\geq$0.438   \\
 V12   &    41437    &    7:38:19.788  & 38:54:40.96 & RRab      &  0.661875  & 52608.6917     &    20.268   &    0.970     &     20.724   &   $\geq$0.850  &   19.792    & indef   \\
 V13   &    24488    &    7:38:16.926  & 38:52:40.20 &RRab      &  0.640267  & 52609.005      &    20.304   &    0.885     &     20.806   & $\geq$1.023    &   19.676    & indef \\
 V14   &    24717    &    7:37:58.391  & 38:52:42.30 & RRab      &  0.741325  & 52610.0986     &    20.238   &    0.962     &     20.668   &   $\geq$0.850   &   19.610    &  $\geq$0.176  \\
 V15   &    32575    &    7:38:13.662  & 38:53:30.62 & RRab      &  0.640006  & 52608.0384     &    20.283   &    0.858     &     20.790   &  $\geq$0.579    &   19.672    & indef   \\
 V16   &    37246    &    7:38:12.380  & 38:54:03.39 & RRab      &  0.666080  & 53032.4811     &    20.295   &    1.124     &     20.669   &   $\geq$1.25   &   19.712    & indef   \\
 V17   &    41439    &    7:38:17.740  & 38:54:40.95 & RRab      &  0.649043  & 52609.060      &    20.319   &    1.134     &     20.764   &  $\geq$1.088    & 19.687    & indef   \\
 V18   &    41939    &    7:38:07.165  & 38:54:46.81 & Cep~II    &  1.578669  & 53032.481      &    18.873   &    0.750     &     19.414   &  $\geq$0.907    &   18.234    &  $\geq$0.492 \\
 V19   &    20251    &    7:37:59.022  & 38:52:15.01 & RRab      &  0.70260   & 52611.1197     &    20.284   &    0.952     &     20.758   &   $\geq$0.387   &   19.735    & $\geq$0.532 \\
 V20   &    33657    &    7:38:05.958  & 38:53:38.19 & LPV     &  52.95     &  52650.0       &    17.386   &  $\geq$0.278 &     18.692   &  $\geq$0.351   &   15.925    & indef   \\
 V21   &    31470    &    7:38:03.576  & 38:53:23.70 & RRab   &  0.686094  & 53032.5260     &    20.202   &    1.074     &     20.676   &  $\geq$1.164   &   19.637    & indef   \\
 V22   &    25180    &    7:38:17.651  & 38:52:44.35 & RRab   &  0.576645  & 53032.4912     &    20.363   &    1.109     &     20.759   &  $\geq$1.304    &   19.847    & $\geq$0.392   \\
 V23   &    38186    &    7:38:10.696  & 38:54:10.73 & RRab   &  0.62650   & 52610.1163     &    20.368   &    1.029     &     20.792   &  $\geq$1.200   &   19.758    &   $\geq$0.212  \\
 V24   &    25237    &    7:37:55.611  & 38:52:45.71 & RRab   &  0.652695  & 52617.1470     &    20.365   &    0.886     &     20.817   &  $\geq$0.750   &   19.747    & indef   \\
 V25   &    32709    &    7:38:03.277  & 38:53:31.71 & RRab   &  0.636274  & 52608.450      &    20.334   &    1.076     &     20.864   &   $\geq$0.800  &   19.746    & $\geq$0.187    \\
 V26   &     7326    &    7:38:02.200  & 38:52:03.20 & RRab    &  0.66492   & 52610.0787     &    20.178   &    0.923     &     20.633   &  1.085    &   19.569    & indef \\
 V27   &     12101   &    7:38:09.776  & 38:51:09.04 & RRc   &   0.34519 &  53033.570     &     20.407  & $\geq$0.275     &      20.782  &    $\geq$0.235   &    19.926   &  indef  \\
 V28   &    36163    &    7:37:51.813  & 38:53:56.14 &  RRab   &  0.64560   & 52609.617      &    20.350   & $\geq$0.675  &   ...   &  ...    &   19.739& $\geq$0.133    \\
 V29   &    25528    &    7:38:03.261  & 38:52:46.94 &  RRab   &  0.725560  & 53032.4811     &    20.278   &    0.777     &     20.664   &   $\geq$0.864   &   19.626    &    $\geq$0.175 \\
 V30   &    30275    &    7:38:06.094  & 38:53:16.18 &  RRab   &  0.584618  & 53032.5910     &    20.322   &    1.065     &     20.513   &  1.343    &   19.745    &  $\geq$0.419  \\
 V31   &  5001910    &    7:38:21.244  & 38:50:22.92 & RRc   &  0.38753   & 52608.89150    &    20.344   &    0.493     &     20.699   &  0.589    &   19.853    & $\geq$0.216    \\
 V32   &    34111    &    7:38:06.700  & 38:53:40.79 &  RRab   &  0.64224   & 52608.3298     &    20.147   &  $\geq$0.680 &     20.610   &  $\geq$0.830   &   19.570    & indef   \\
 V33   &    23396    &     7:38:12.269 &  38:52:33.92 & RRc   &  0.41117   & 52608.7635     &    20.323   &    0.474     &     20.685   &  0.535    &   19.824    & 0.09   \\
 V34   &    44725    &     7:38:10.337 &  38:55:29.06 & RRc   &  0.403547  & 53053.5391     &    20.364   &    0.482     &     20.750   &  0.595    &   19.845    & indef  \\
 V35   &    27712    &     7:38:12.009 &  38:52:59.86 &   RRab   &  0.677207  & 53053.58012    &    20.265   &    0.867     &     20.695   &  1.097    &   19.614    & indef    \\
 V36   &    33246    &     7:38:10.301 &  38:53:35.31 & RRab   &  0.64822   & 52991.6243     &    20.299   &    0.773     &     20.644   &  $\geq$0.907    &   19.676    & 0.227  \\
 V37   &    29181    &     7:38:11.197 &  38:53:09.06 & RRab   &  0.6610265 & 52991.6340     &    20.276   &    $\geq$1.056     &     20.582   &  1.203    &   19.714    & indef   \\
 V38   &    18676    &     7:38:08.132 &  38:52:03.94 & RRc   &  0.368980  & 52911.6340     &    20.248   &    0.457    &     20.583   &  0.571   &   19.791    & $\geq$ 0.286  \\
 V39   &    10671    &     7:38:10.715 &  38:50:51.10 & RRd    &  0.40703   & 52609.070      &    20.344   &    0.474     &     20.571   &  $\geq$0.409    &   19.909    & $\geq$ 0.265   \\
 V40   &    25629    &     7:38:13.089 &  38:52:47.21 &  RRab   &  0.5756470 & 50770.1022     &    20.217   &    1.074     &     20.605   &  1.359    &   19.631    & 0.922  \\
 V41   &    31405    &     7:38:07.200 &  38:53:23.10 &  RRc   &  0.396460  & 53032.51155    &    20.334   &    0.430     &     20.751   & $\geq$ 0.560     &   19.834    & 0.233    \\
 V42   &   25357     &     7:38:08.719 &  38:52:45.74 &  RRab     &0.610078  &52806.7635       &    20.293   &    0.758     &     20.036   &  $\geq$0.882   &   19.689    & $\geq$0.704  \\
 V43   &   23674     &     7:38:07.564 &  38:52:35.13 & RRc     &0.379955  &52610.154        &    20.222   &    0.147     &     ...   &  ...    &   19.814  & indef  \\
 V44   &   24629     &     7:38:09.672 &  38:52:41.32 & NC     &0.87583   &52906.7493       &    17.393   &    0.174     &     18.681   &  0.243    &   15.898    & indef   \\
 V45   &   26853     &     7:38:06.530 &  38:52:54.84 & RRc     &0.31394   &52610.12         &    20.387   &    0.195     &     20.188   &  $\geq$0.602     &   ...    & ...   \\
 V46   &   26160     &     7:38:06.448 &  38:52:50.76 & RRab     &0.774047  &50771.35         &    20.217   &$\geq$0.622   &     20.454   &   $\geq$0.442    &   19.679    & $\geq$0.135   \\
 V47   &   15945     &     7:38:08.600 &  38:53:14.68 &  RRab        &0.65684   &52617.0368       &    20.312   &$\geq$0.622   &     20.540   &  $\geq$0.296    &   19.748    & indef    \\
 V48   &   24628     &     7:38:10.035 &  38:52:41.15 & RRc     &0.374897  &50771.27149      &    20.214   &    0.476     &     20.367   &   $\geq$0.493  &   19.762    & $\geq$0.250  \\
 V49   &   30239     &     7:38:08.131 &  38:53:15.87 & RRc        &0.327090  &50771.23356      &    20.364   &    0.510     &     20.157   &  $\geq$0.301    &   19.959    & 0.361  \\
 V50   &   16182     &     7:38:07.980 &  38:53:16.22 & RRab      &0.698385  &50770.68         &    20.191   &$\geq$0.653   &     20.254   &  $\geq$0.617   &   19.522    &$\geq$0.358 \\
 V51   &   27292     &    7:38:06.188  & 38:52:57.50 & RRc     &0.348446  & 49493.477        &    20.195   &    0.530     &     ...   &  ...    &   19.781    & 0.342    \\
 V52   &   29702     &    7:38:09.594  & 38:53:12.36 & RRc     &0.332284  & 52609.0579       &    20.427   &    0.410     &     20.472   &  $\geq$0.469    &   19.856    & $\geq$0.273   \\
 V53   &   30464     &   7:38:07.782   &38:53:17.29  &RRab        &0.658467  & 50770.7191       &    20.007   &    0.805     &     20.347   &  $\geq$0.889    &   ...    & ...   \\
 V54   &   13009     &  7:38:10.532   &38:52:50.64 &RRc     &0.43164   & 52609.11         &    19.410   &    0.260     &     19.622   &  0.146     &   18.860    & $\geq$0.081   \\
 V55   &   27613     &   7:38:05.994   &38:52:59.62  &RRc     &0.387675  & 53032.500        &    20.311   &    0.400     &     20.599   &  0.447    &   19.832    & $\geq$0.323    \\
 V56   &   21790     &   7:38:07.987   &38:52:24.65  &RRc     &0.305795  & 49495.320        &    20.286   &    0.287     &     ...   &  ...    &   19.864    & indef    \\
 V57   &   24846     &   7:38:06.004   &38:52:42.72  &RRab     &0.736379  & 51622.095        &    19.948   &    0.741     &     20.511   &  $\geq$0.713    &   19.316    & 0.575    \\
 V58   &   27780     &   7:38:10.644   &38:53:00.3  &RRc     &0.388238  & 52617.060        &    20.340   &    0.472     &     20.560   &  0.546    &   19.863    & $\geq$0.290   \\
 V59   &   29251     &   7:38:10.426   &38:53:09.50  &RRab     &0.803492  & 53033.515        &    20.233   &    0.527     &     20.562   &  0.593    &   19.559    & $\geq$0.358   \\
 V60   &   30716     &   7:38:06.659   &38:53:18.83  &RRc     &0.390496  & 52991.640        &    20.299   &    0.509     &     20.577   &  0.523    &   19.756    & $\geq$0.242   \\
 V61   &    8720     &  7:38:08.588    &38:52:16.08  &SX~Phe      &0.05265   & 52608.9791       &    23.069   &    0.713     &     22.107   &  indef   &   22.655    & indef    \\
 V62   &   17882     &   7:38:08.324   &38:53:31.44  &RRab      &0.592503  & 52609.272        &    20.328   &    1.025     &     20.398   &  $\geq$1.104    &   19.786    & $\geq$0.549    \\
 V63   &   30364     &   7:38:10.979   &38:53:16.44  &RRc     &0.344915  & 52991.614        &    20.306   &    0.519     &     20.592   &  0.595    &   19.872    & 0.350   \\
 V64   &   20512     &   7:38:10.293   &38:52:16.34  &RRab     &0.780009  & 49494.54495      &    20.314   &    0.456     &     20.526   &  $\geq$0.419    &   19.633    & indef   \\
 V65   &   22441     &   7:38:11.321   &38:52:28.18  &RRc   &0.32278   & 52610.1154       &    20.109   &    0.230     &     20.691   &  0.341    &   19.453    & indef    \\
 V66   &   29417     &   7:38:11.600   &38:53:10.46  & RRc     &0.386754  & 52611.143        &    20.340   &    0.443     &     20.523   &  0.593    &   19.863    & $\geq$0.261  \\
 V67   &   31497     &   7:38:10.970   &38:53:23.47  &RRc     &0.33989   & 52905.7417       &    20.364   &    0.453     &     20.672   &  0.557    &   19.922    & indef   \\
 V68   &   25927     &   7:38:12.147   &38:52:48.97  &RRc     &0.364675  & 50768.5526       &    20.335   &    0.499     &     20.744   &  0.6    &   19.872    &   0.315  \\
 V69   &   33596     &   7:38:10.641   &38:53:37.73  & RRc     &0.340932  & 53053.604        &    20.430   &    0.500     &     20.927   &  $\geq$0.306    &   19.954    & $\geq$0.062   \\
 V70   &   19518     &   7:38:08.468   &38:53:46.63  &RRc     &0.4155    & 52991.62         &    20.143   &    0.339     &     20.741   &  0.432    &   19.329    & $\geq$0.346   \\
 V71   &   17581     &   7:38:07.334   &38:51:56.28 &SX~Phe      &0.049197  & 52608.976        &    23.066   &    0.418     &     23.459   &  indef    &   22.730    & $\geq$0.166    \\
 V72   &   19747     &    7:38:09.923  & 38:53:49.52 & RRc     &0.414626  & 52991.63404      &    20.376   &    0.455     &     20.772   &  0.577    &   19.828    & $\geq$0.088  \\
 V73   &   18388     & 7:38:11.521   &38:52:01.76 &SX~Phe     &0.0528674 & 53053.57076      &    22.967   &    0.383     &     23.280   &  $\geq$0.325   &   22.505    & indef   \\
 V74   &   25759     &   7:38:13.777   &38:52:47.94  &RRc     &0.305679  & 50771.07686      &    20.414   &    0.200     &     20.723   &  0.225    &   ...    & ...  \\
 V75   &   20587     &   7:38:03.540   &38:52:16.88  & RRc     &0.3236    & 52991.6243       &    20.352   &    0.443     &     20.672   &  0.570    &   19.941    & $\geq$0.228  \\
 V76   &   18026     &   7:38:12.110   &38:51:59.23  &RRc     &0.44105   & 52608.650        &    20.301   &    0.474     &     20.544   &  0.531    &   19.795    & 0.278   \\
 V77   &   19320     &   7:38:13.139   &38:52:08.23  &RRc     &0.38113   & 53032.4930       &    20.379   &    0.489     &     20.750   &  $\geq$0.526    &   19.892    &  $\geq$0.151  \\
 V78   &   31821     &   7:38:13.752   &38:53:25.53  &SX~Phe     &0.0411654 & 52905.7415       &    23.027   &    0.273     &     23.394   &  $\geq$0.287    &   22.690    & $\geq$0.142   \\
 V79   &   18137     &   7:38:13.194   &38:52:00.02  &SX~Phe     &0.049438  & 52610.058        &    22.801   &    0.767     &     22.846   &  0.739    &   22.443    & $\geq$0.453   \\
 V80   &   16192     &  7:38:01.810   &38:53:16.64  &RRab        &0.6161    & 53053.595        &    20.648   &    0.604     &     21.336   &  $\geq$0.838    &   ...    & ...   \\
 V81   &   36489     &  7:38:12.102   &38:53:57.42  &SX~Phe     &0.0628808 & 53053.5807       &    22.651   &    0.424     &     22.975   &  $\geq$0.471    &   22.107    & indef   \\
 V82   &   33557     &    7:38:13.847  & 38:53:37.34 & RRc     &0.34728   & 53053.58012      &    20.318   &    0.303     &     20.658   &  0.376    &   19.912    & $\geq$0.155  \\
 V83   &   14335     & 7:38:08.535   &38:51:31.40  & SX~Phe(?)   &0.14033   & 52608.5445       &    22.548   &    0.199     &     22.846   &  0.075    &   22.183    & indef    \\
 V84   &   29591     &   7:38:01.459   &38:53:11.99  & RRc     &0.3271    & 52609.052        &    20.323   &    0.416     &     20.722   &  $\geq$0.225    &   19.921    &  $\geq$0.163  \\
 V85   &   38108     &   7:38:01.849   &38:54:10.44  &SX~Phe     &0.0499945 & 52608.97575      &    22.906   &    0.413     &     23.313   &  $\geq$0.498    &   22.587    & $\geq$0.218   \\
 V86   &   30259     &   7:38:19.618   &38:53:15.74  &LPV       &49.6      & 53053.591        &    17.466   &    0.289     &     18.825   &  0.437    &   15.919    & indef  \\
 V87   &   13811     &   7:38:18.402   &38:51:26.32  &Bin      &0.233955  & 52617.14485      &    23.157   &    1.200     &     23.540   &  0.263    &   22.726    & 0.536   \\
 V88   &    7517     &   7:38:10.573   &38:49:53.37  &SX~Phe      &0.05372   & 52611.1386       &    22.879   &    0.281     &     23.126   &  $\geq$0.193   &   22.473    & 0.168  \\
 V89   &   43459     &   7:38:19.777   &38:55:07.35  &RRc     &0.28972   & 52608.8022       &    20.409   &    0.079     &     20.701   &  0.064    &   20.014    & indef    \\
 V90   & 5009772     &   7:38:23.136   &38:54:11.64  &RRc     &0.39742   & 52608.71826      &    20.392   &    0.457     &     20.765   &  $\geq$0.490    &   19.894    & indef  \\
 V91   &   38530     &   7:37:52.824   &38:54:14.42  &RRc     &0.3893    & 52609.990        &    20.379   &    0.475     &     ...   &  ...    &   19.846    & indef   \\
 V92   &      347    &   7:38:51.641   &38:53:49.61  & NC     & 5.63       &52608.22        &    19.548   &    0.112     &     ...   &  ...    &   18.162    & indef   \\
 V93   &     2265    &   7:37:49.886   &38:44:59.89  &SX~Phe    & 0.05659    &52608.5583      &    22.923   &    0.216     &     ...   &  ...    &   22.463    & $\geq$0.100   \\
 V94   &  5015266    &   7:38:26.106   &38:58:10.17  & RRab   & 0.71332    &52610.050       &    20.318   &    0.590     &     ...   &  ...    &   19.659    & indef   \\
 V95   &     3462    &   7:37:29.362   &38:54:33.92  &$\delta$Scuti& 0.1483 &52608.8198      &    19.228   &    0.064     &     ...   &  ...    &   18.441    & 0.070   \\
 V96   &     2513    &   7:37:39.605   &38:51:35.81  &Bin     & 0.3597     &52608.912       &    19.085   &    0.887     &     ...   &  ...    &   17.629    & indef    \\
 V97   &     3255    &   7:38:18.406   &38:49:51.32 &SX~Phe    & 0.05017    &52608.9515      &    22.690   &    0.327     &     ...   &  ...    &   22.341    & $\geq$0.162   \\
 V98   &     2254    &   7:37:38.956   &38:50:22.02  &   NC     & 0.2296     &52610.0866      &    22.553   &    0.604     &     ...   &  ...    &   20.221    & 0.285   \\
 V99   &  5001168    &   7:38:28.628   &38:49:15.4   &Bin   & 0.2015     &526609.633      &    18.526   &    0.127     &     ...   &  ...    &   16.461    & indef   \\
 V100  &     4804    &   7:37:52.712   &38:48:11.55 &$\delta$Scuti& 0.13090 &52608.940       &    18.335   &    0.264     &     ...   &  ...    &   17.423    & indef    \\
 V101  &     3528    &   7:38:14.881   &38:50:24.74 &SX~Phe   & 0.05038    &52608.995       &    22.714   &    0.459     &     ...   &  ...    &   22.080    & indef   \\
\enddata
\label{tabellone}
\end{deluxetable}
\begin{table}[h]
\centering
\caption{Distances to NGC~2419 obtained using diferent methods based on the cluster RR Lyrae stars.}
\vspace{0.5 cm}
\begin{tabular}{ccc}
\hline
\hline
{\rm (m-M)$_0$} & {\rm error}  & {\rm method}\\
{\rm (mag)} & (mag)  & \\
\hline
 19.65  & $\pm$0.09    &  M$_V$-[Fe/H]  \\
 19.63  & $\pm$0.11    &  FOBE  \\
 19.70  & $\pm$0.09    &  PW  \\
 19.67  & $\pm$0.10    &  PLI  \\
 19.83  & $\pm$0.11    &  FIT-V34  \\
 19.73  & $\pm$0.07    &  FIT-V75 \\
 19.81  & $\pm$0.08    &  FIT-V3  \\
\hline
\end{tabular}
\label{moduli}
\end{table}

\end{document}